\documentclass[epj]{svjour}

\usepackage{amsmath}
\usepackage{amssymb}
\usepackage[tight]{units}
\usepackage{textcomp}
\usepackage{array}
\usepackage{color,graphicx}
\usepackage[numbers]{natbib}
\usepackage[bookmarks=false]{hyperref}

\newcommand{\dd}{\mathrm{d}}

\begin{document}

\title{Results from 
\texorpdfstring{\unit[730]{kg\,days}}{730 kg days} of the CRESST-II Dark Matter Search}

\author{
G.~Angloher\inst{1}\and
M.~Bauer\inst{2}\and
I.~Bavykina\inst{1}\and
A.~Bento\inst{1,5} \and
C.~Bucci\inst{3}\and
C.~Ciemniak\inst{4}\and
G.~Deuter\inst{2}\and
F.~von~Feilitzsch\inst{4}\and
D.~Hauff\inst{1}\and
P.~Huff\inst{1}\and
C.~Isaila\inst{4}\and
J.~Jochum\inst{2}\and
M.~Kiefer\inst{1}\and
M.~Kimmerle\inst{2}\and
J.-C.~Lanfranchi\inst{4}\and 
F.~Petricca\inst{1}\and
S.~Pfister\inst{4}\and
W.~Potzel\inst{4}\and
F.~Pr\"obst\inst{1}\thanks{corresponding author: proebst@mpp.mpg.de}\and
F.~Reindl\inst{1}\and
S.~Roth\inst{4}\and
K.~Rottler\inst{2}\and 
C.~Sailer\inst{2}\and
K.~Sch\"affner\inst{1}\and
J.~Schmaler\inst{1}\thanks{corresponding author: jschmale@mpp.mpg.de}\and
S.~Scholl\inst{2}\and
W.~Seidel\inst{1}\and
M.~v.~Sivers\inst{4}\and
L.~Stodolsky\inst{1}\and
C.~Strandhagen\inst{2}\and
R.~Strau\ss\inst{4}\and
A.~Tanzke\inst{1}\and
I.~Usherov\inst{2}\and
S.~Wawoczny\inst{4}\and
M.~Willers\inst{4}\and
A.~Z\"oller\inst{4}
}


\institute{
Max-Planck-Institut f\"ur Physik, D-80805 M\"unchen, Germany \and
Eberhard-Karls-Universit\"at T\"ubingen, D-72076 T\"ubingen, Germany \and
INFN, Laboratori Nazionali del Gran Sasso, I-67010 Assergi, Italy \and
Physik-Department E15, Technische Universit\"at M\"unchen, D-85747 Garching, Germany \and
Departamento de Fisica, Universidade de Coimbra, P3004 516 Coimbra, Portugal
}

\date{ }

\abstract{
The CRESST-II cryogenic Dark Matter search, aiming at detection of
WIMPs via elastic scattering off nuclei in CaWO$_4$ crystals, completed 730\,kg\,days of data taking in 2011. We present the data
collected with eight detector modules, each with a two-channel
readout; one for a phonon signal and the other for coincidently produced
scintillation light. The former provides a precise measure of the
energy deposited by an interaction, and the ratio of scintillation
light to deposited energy can be used to discriminate different
types of interacting particles and thus to distinguish possible
signal events from the dominant backgrounds.\\
Sixty-seven events are found in the acceptance region where  a WIMP
signal in the form of low energy nuclear recoils would be expected. We
estimate background contributions to this observation from four sources: 1)
``leakage'' from the $e/\gamma$-band 2)``leakage'' from the 
  $\alpha$-particle band 3)
neutrons and 4) $^{206}$Pb recoils from $^{210}$Po decay.  Using a maximum likelihood
analysis, we find, at a high statistical significance, that these
sources alone are not sufficient to explain the data. The addition
of a signal due to scattering of relatively light WIMPs could
account for this discrepancy, and we determine the associated WIMP
parameters.  
}

\maketitle

\section{Introduction}

The nature of Dark Matter remains one of the outstanding questions
of present-day physics. There is convincing evidence for its
existence on all astrophysical scales and many theories predict
particle candidates that may be able to explain its composition.
However, in spite of numerous attempts, Dark Matter particles have 
not been unambiguously detected so far. 

Several experiments currently aim for direct detection of Dark
Matter, mostly focusing on a particular class of particles, the
so-called WIMPs (Weakly Interacting Massive Particles). WIMPs
nowadays are among the most investigated and best motivated
candidates to explain Dark Matter. If they exist, they could be
present in our galaxy in the form of a halo, constituting the
majority of the galactic mass. Rare interactions with ordinary
matter would then possibly be detectable in earthbound experiments.

One such project is CRESST-II (Cryogenic Rare Event Search with
Superconducting Thermometers), located at the Laboratori Nazionali
del Gran Sasso in Italy. In this experiment we aim for  detection
of the WIMPs via their scattering off nuclei. The main challenges of
this kind of measurement are to detect the tiny amounts of recoil energy
transferred to the nucleus
($\mathcal{O}(\unit[10]{keV})$), and to achieve sufficient
background suppression to be sensitive to the extremely low rate
of anticipated interactions (not more than a few tens of events per
kilogram and year). 

To meet these requirements, CRESST uses cryogenic detectors
in a low-background environment. Furthermore, the scintillating target material CaWO$_4$ allows for a discrimination of the type of
interacting particle. In this way, potential rare WIMP interactions
can be distinguished from events which were induced by the dominant
radioactive backgrounds. 

In this article, we report on the latest data obtained with the CRESST setup. They comprise a total net exposure of \unit[730]{kg\,d}, collected between 2009 and 2011. The structure of the paper is as follows: Section~\ref{sec:experimentalsetup} starts with a brief introduction to the experimental setup. Section~\ref{sec:latestrun} gives the new aspects of the present run and summarizes the observations. The choice of an appropriate acceptance region and the possible backgrounds relevant in this region are discussed. In Section~\ref{sec:qualitativebckdiscussion}, we give a qualitative description and estimation of the backgrounds.  Section~\ref{sec:likelihoodanalysis} then describes the quantitative treatment in the framework of a maximum likelihood analysis. The results of this analysis are then discussed in Section~\ref{sec:resultsanddiscussion}. Finally, Section~\ref{sec:futuredevelopments} gives the outlook for future runs of the experiment.

\section{Experimental Setup} \label{sec:experimentalsetup}

A detailed description of the CRESST-II setup was the subject of
earlier publications
\cite{Angloher2009_run30,Angloher2005_cresstIIproof}. Here, we
restrict ourselves to a few key aspects relevant for the discussion
of the new results.

\subsection{Target Material}

As a target for WIMP interactions, CRESST uses scintillating
CaWO$_4$ crystals. They have a cylindrical shape (\unit[40]{mm} in
diameter and height) and weigh about \unit[300]{g} each. The 
current experimental setup can accommodate up to 33 of these
crystals, constituting a maximum target mass of about
\unit[10]{kg}.

\begin{figure}
 \centering
 \includegraphics[width=1.0\linewidth]{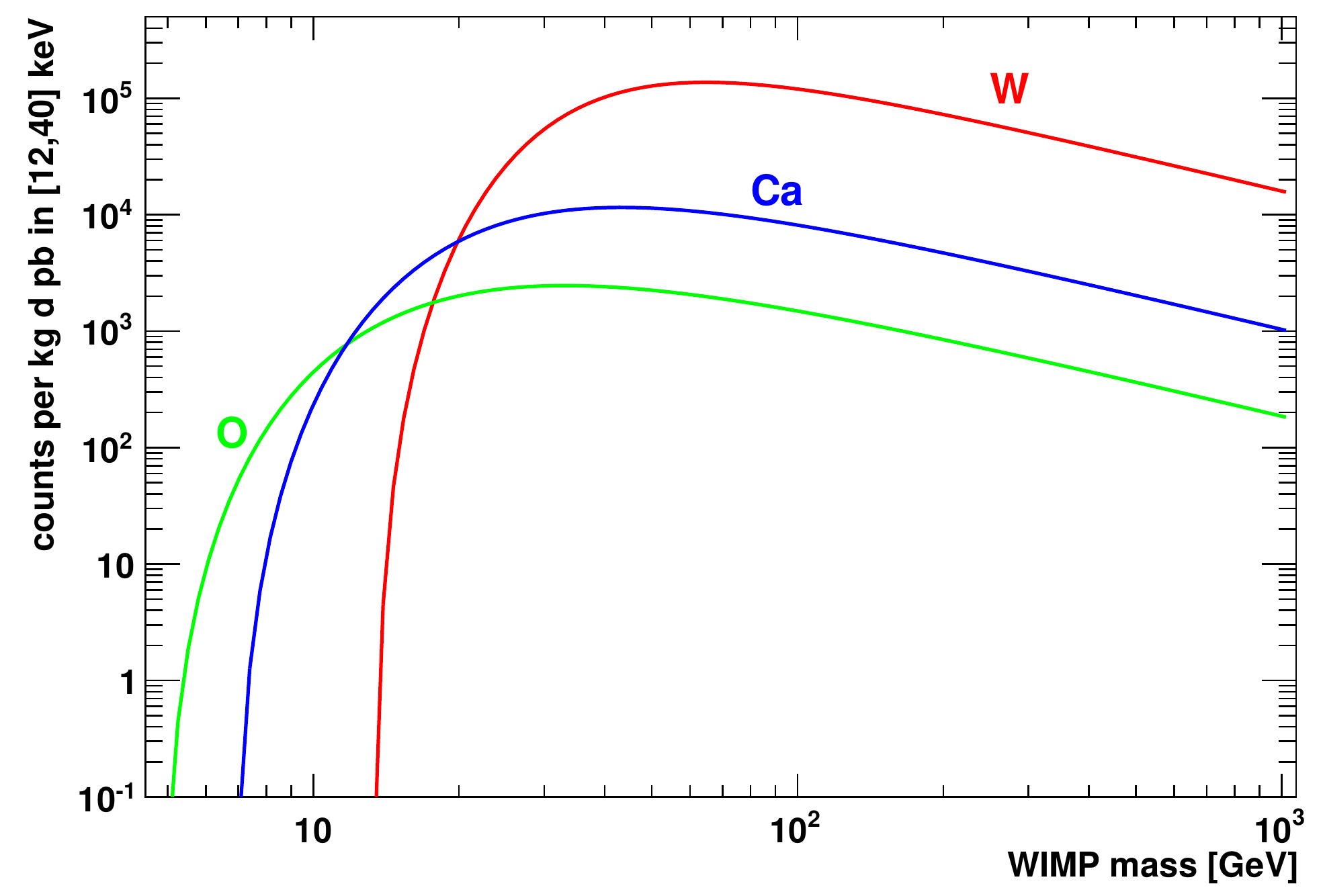}
 \caption{Contributions of the three types of nuclei present in a
CaWO$_4$ target to the total rate of WIMP interactions, as a
function of the WIMP mass and for a cross-section of \unit[1]{pb},
assuming coherent $\sim A^2$ interactions.
The calculation assumes a detector with a sensitive energy range
between 12 and \unit[40]{keV}.}
 \label{fig:targetnuclei}
\end{figure}

Under the usual assumption of coherent WIMP scatterings off nuclei, the scattering cross section 
contains an $A^2$ enhancement (with the mass number $A$ of the
target nucleus). In this case one expects that the
total scattering rate in CaWO$_4$ is dominated by interactions with the heavy
tungsten nuclei. However, due to kinematics a heavier nucleus tends
to receive a smaller recoil energy, and in a detector with a finite
energy threshold the other constituents can also become relevant
despite the coherence effect. To illustrate this point,
Fig.~\ref{fig:targetnuclei} shows, as a function of the WIMP mass, the contributions of the three elements in CaWO$_4$ to the total rate of WIMP interactions in the energy
interval 12 to \unit[40]{keV}, the range  typical of the CRESST
detectors. 
 
For low WIMP masses, up to about \unit[12]{GeV}, the scatterings
off tungsten  are completely below threshold, and oxygen and
calcium recoils give the only possible Dark Matter signal. On the
other hand, above WIMP masses of about \unit[30]{GeV}, tungsten
completely dominates. When looking for possible WIMP interactions
we therefore consider recoils of all three types of nuclei in
CaWO$_4$, in order to be sensitive to the largest possible range of
WIMP masses.

\subsection{Phonon and Light Detectors}

In order to be able to detect the low energy nuclear recoils the
 target crystals are operated as cryogenic calorimeters at temperatures
of about \unit[10]{mK}. The energy deposited by an interacting
particle is mainly converted into phonons, which are then
detected with a transition edge sensor (TES).  We thus denote the
target crystals with their TES as \emph{phonon detectors}. 

The TES is a thin tungsten film evaporated onto the crystal, with
the temperature  stabilized in the transition from the normal to
the superconducting state. The tiny change of the film temperature
($\mathcal{O}$(\textmu K)) induced by the absorption of the 
phonons  leads to a measurable change in resistance. This signal is
read out by SQUID-based electronics. The amplitude of
the signal is a precise measure of the deposited energy. After the
interaction, the crystal temperature relaxes back to the
equilibrium state via a weak thermal coupling to the heat bath.  

In addition to the phonon signal, a small fraction of the energy
deposited in the target crystal is converted into scintillation
light. Each crystal is paired with a separate cryogenic light
detector in order to detect this light signal. Most of the light
detectors are made from a silicon-on-sapphire wafer (a sapphire
wafer of \unit[40]{mm} diameter and \unit[0.46]{mm} thickness, with
a \unit[1]{\textmu m} silicon layer on one side, which acts as
photon absorber). As an alternative, some light detectors consist
of pure silicon wafers of the same size. Similar to the target
crystals, each light detector is equipped with a tungsten
transition edge sensor to determine the energy deposited by the
absorption of scintillation photons.

\begin{figure}
  \centering
  \includegraphics[width=\linewidth]{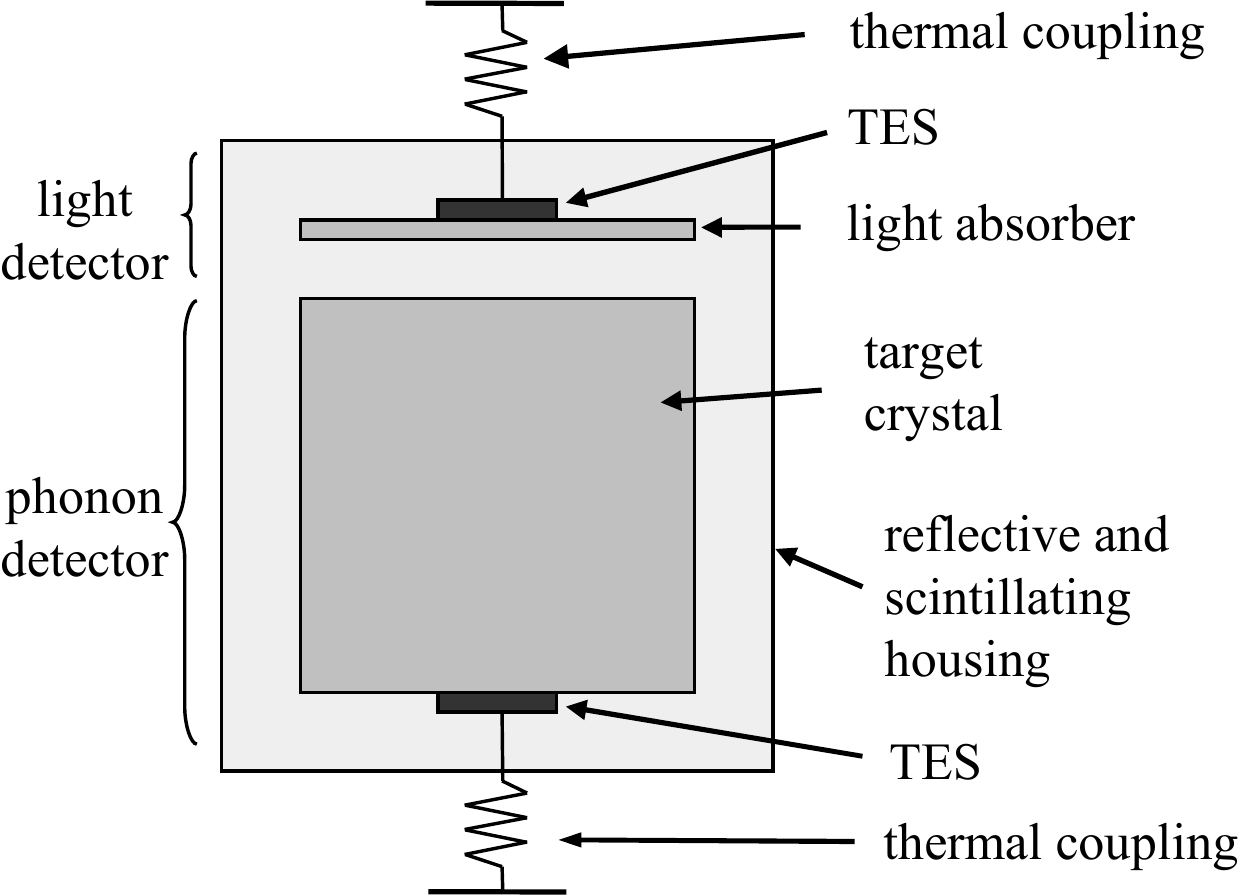}
  \caption{Schematic drawing of a CRESST detector module,
consisting of the target crystal and an independent light detector.
Both are read out by transition edge sensors (TES) and are enclosed
in a common reflective and scintillating housing.}
  \label{fig:detectormodule}
\end{figure}

A crystal and the corresponding light detector form a so-called
\emph{detector module} as shown in Fig.~\ref{fig:detectormodule}.
Both detectors of a module are held by thin, silver-coated bronze
clamps and are enclosed in a common, highly light reflective
housing in order to collect as much scintillation light as
possible. The reflector is a polymeric foil which also 
scintillates. This will be discussed in detail in
Section~\ref{sec:scintillatinghousing}.

\subsection{Quenching Factors and Background Discrimination} \label{sec:activebackgrounddiscrimination}

For each particle interaction, a detector module yields two
coincident signals (one from the phonon and one from the light
detector). While the phonon channel provides a sensitive
measurement of the total energy deposition in the target
(approximately independent of the type of interacting particle),
the light signal can be used to discriminate different types of
interactions. To this end, we define the \emph{light yield} of an
event as the ratio of energy measured with the light detector
divided by the energy measured with the phonon detector. We
normalize the energy scale of the light channel such that
\unit[122]{keV} $\gamma$'s from a $^{57}$Co calibration source 
have a light yield of unity. With this normalization electron recoils induced
either by $\beta$ sources  or by  gamma interactions, generally
have a light yield of about 1. On the other hand,  $\alpha$'s 
and nuclear recoils are found to have a lower light yield. We quantify this
reduction by assigning a \emph{quenching factor} (QF) to each type
of interaction. The QF describes the light output expressed as a percentage of
the light output for a $\gamma$ of the same deposited energy. 

Some quenching factors can be directly determined from  CRESST
data. For example, neutrons detectably scatter mainly off the oxygen nuclei in CaWO$_4$. The QF for oxygen can thus be
determined from a neutron calibration run which took place during
the  data taking discussed here. The result is
\begin{displaymath}
 QF_{\text{ O }} = (10.4 \pm 0.5)\,\%.
\end{displaymath}
Moreover, the quenching factor for low energy $\alpha$'s can be
found to be about \unit[22]{\%}, using  $\alpha$-events  in the
current data set. Similarly, the value for lead  can be inferred to
be around \unit[1.4]{\%}. Both measurements will be discussed
below. 

Other types of interactions (in particular calcium and tungsten
recoils) are not observed with sufficient statistics in CRESST, and
their quenching factors must be determined in dedicated experiments
\cite{Huff2011_QuenchingFactors}:
\begin{align*}
 QF_{\text{Ca}} &= \left( 6.38 ^{+0.62}_{-0.65} \right) \% \\
 QF_{\text{W}} &= \left( 3.91 ^{+0.48}_{-0.43} \right) \%.
\end{align*}
Corresponding to these different values, there will be
characteristic ``bands'' for the different particles or recoils in
the light yield-energy plane. This allows for an excellent
discrimination between potential signal events (expected to show up
as nuclear recoils)  and the dominant radioactive backgrounds
(mainly $e/\gamma$-events). 

Furthermore, it is even partially possible to  determine
which  type of nucleus is recoiling.
Such a  discrimination is possible to the extent to which the
different nuclear recoil bands in the light yield-energy plane can
be separated within the resolution of the light channel.  This then
allows a study of potential WIMP interactions  with different
target nuclei, in parallel in the same setup. Such a possibility can be particularly relevant for the verification of a
positive WIMP signal, and is a distinctive feature of CRESST.

\subsection{Scintillating Housing} \label{sec:scintillatinghousing}

\begin{figure}
  \centering
  \includegraphics[width=\linewidth]{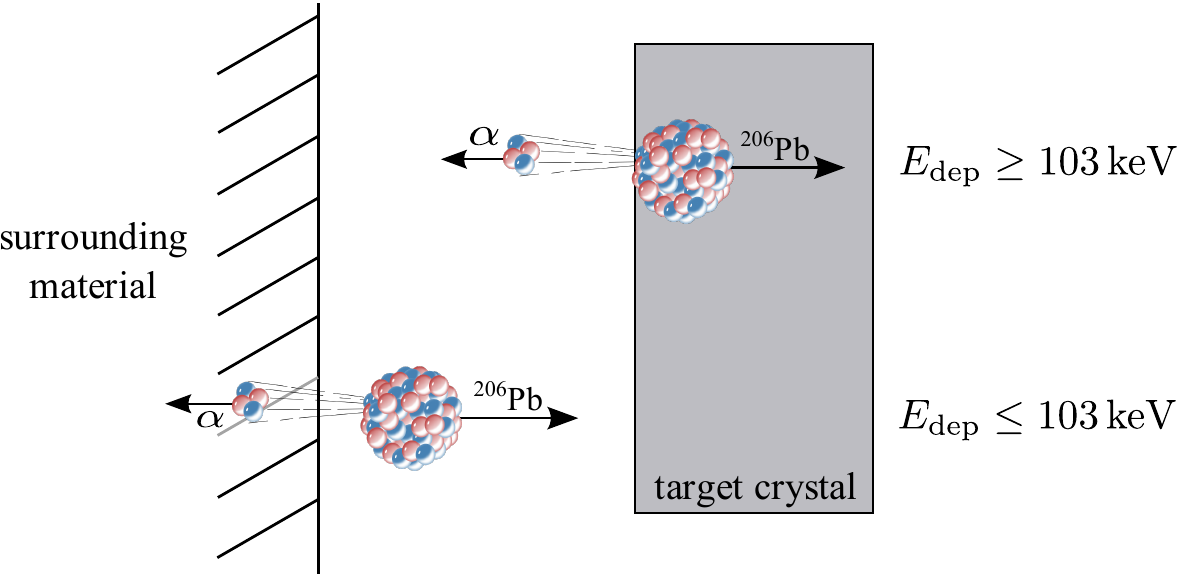}
  \caption{Illustration of background events due to surface
contaminations with $^{210}$Po.}
  \label{fig:surfacealpha}
\end{figure}

As mentioned above, the housing of the detector modules consists  
mainly of a reflecting and scintillating polymeric foil. Making all
surfaces in the vicinity of the detectors scintillating is 
important in discriminating background events due to  contamination
of surfaces with $\alpha$-emitters. The basic mechanism of this
background is illustrated in Fig.~\ref{fig:surfacealpha}.

The most important isotope in this context is $^{210}$Po, a decay product of the gas $^{222}$Rn. It can be present on or
implanted in the surfaces of the detectors and surrounding
material. The $^{210}$Po nuclei decay to $^{206}$Pb, giving a
\unit[5.3]{MeV} $\alpha$-particle and  a \unit[103]{keV} recoiling lead nucleus.
 It can happen that the lead nucleus hits
the target crystal and deposits its energy there,  while the
$\alpha$-particle escapes. Due to its low quenching factor, the
lead nucleus can often not be distinguished from a tungsten recoil 
and thus can mimic a WIMP interaction. 

However, if the
 polonium mother nucleus was located on the surface of the target
crystal or  implanted in it (the upper case in
Fig.~\ref{fig:surfacealpha}),  the full \unit[103]{keV} of the
daughter nucleus plus a possible contribution from the escaping
$\alpha$-particle will be deposited in the target and  the event
will lie above the energy range relevant for the WIMP search. 

 Another situation arises when the polonium atom was implanted in
a surrounding surface. Then the daughter Pb nucleus can lose part
of its energy on the way to the target crystal and appear in the energy
range of interest (the lower case in Fig.~\ref{fig:surfacealpha}).
This possibly  dangerous   background  can be rendered harmless if
the surrounding surfaces are scintillating;  in this case the escaping
$\alpha$-particle produces additional scintillation light when
hitting those surfaces and the event will appear as high-light
event in distinction to the  low-light nuclear recoils. 

Hence the scintillation of the complete surroundings of the target
crystals plays an  important role.
With the scintillating foil used as a module housing, currently the
only non-scintillating surfaces inside the detector modules are the
small clamps which hold the target crystals. In earlier runs,
attempts were made to cover these clamps with scintillating layers
as well, but these layers appeared to give rise to thermal
relaxation events. The current module design therefore avoids any
scintillating (plastic) material in direct contact with the
crystals. The price for this measure, however, is the presence of
several Pb recoil events with energies of \unit[103]{keV} and below
in the data set, as expected from the above discussion. This
background must therefore be taken into account in our analysis.

\subsection{Shielding}

\begin{figure}[t!]
  \centering
  \includegraphics[width=\linewidth]{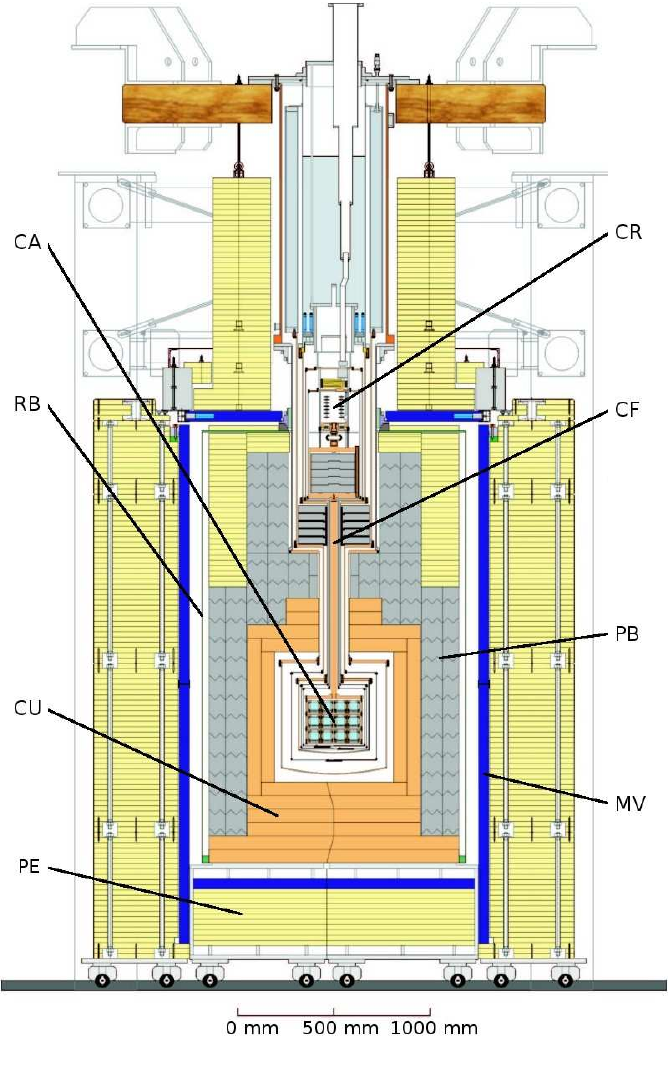}
  \caption{Schematic drawing of the CRESST setup. A cold finger
(CF) links the cryostat (CR) to the experimental volume, where the
detectors are arranged in a common support structure, the so-called
carousel (CA). This volume is surrounded by layers of shielding
from copper (CU), lead (PB), and polyethylene (PE). The copper and
lead shieldings are additionally enclosed in a radon box (RB). An
active muon veto (MV) tags events which are induced by cosmic
radiation.}
  \label{fig:cresstsetup}
\end{figure}

Fig.~\ref{fig:cresstsetup} shows a schematic drawing of the whole
CRESST setup, with the detector modules in the very center. 
 The low temperatures are provided by a $^3$He-$^4$He dilution
refrigerator and transferred to the detectors via a \unit[1.3]{m}
long  copper  cold finger. The detector volume is surrounded by
several layers of shielding against the main types of background
radiation: layers of highly pure copper and lead  shield against
$\gamma$-rays, while polyethylene serves as a moderator for
neutrons. The inner layers of shielding are contained in a gas
tight box to prevent radon from penetrating them. In addition, an active
muon veto using plastic scintillator panels is installed  to tag
muons. The veto surrounds the lead and copper shielding and covers
98.7\,\% of the solid angle around the detectors,  a small hole on
top is necessary to leave space for the cryostat.

\subsection{Data Analysis}

We apply just a few basic quality cuts to the raw data in order to
ensure that only valid events, with well-reconstructed energies in
the phonon and light channel,  are considered for further analysis.
In particular, we require that both the phonon and light detector
of a given module were fully operational and running stably at
their respective operating points at the time of an event. Data
acquisition, readout, and
 the procedures for  monitoring detector stability, trigger
efficiency as well as reconstructing the deposited energy from the
measured pulses are described elsewhere~\cite{Angloher2009_run30}. 

For the final data set, we reject events coincident with a signal
in the muon veto as well as those events with coincident signals in more than one
detector module, since multiple scatterings are excluded for
WIMPs in view of their rare interactions.

An important aspect of the analysis concerns the bands in the light
yield-energy plane. The $e/\gamma$-band is highly populated due to
the relatively high rate of common backgrounds. This allows us to
extract the position and, in particular, the energy-dependent width
of this band directly from the measured data. The observed width is
dominated by the light channel resolution compared to which the resolution of the phonon channel is much
superior. This is understandable in
view of the small fraction of the deposited energy appearing as light.

We extract the resolution of the light channel as a function of
light energy by fitting the $e/\gamma$-band with a Gaussian of
energy dependent center and width. We note that, although the production of
scintillation light is governed by Poisson statistics, the Gaussian
model is a very good approximation in our regions of interest. This
is because  the $e/\gamma$-events produce a sufficiently large
number of photons for the Poisson distribution to be well
approximated by a Gaussian. On the other hand, for the quenched
bands with low light yields, the Gaussian baseline noise  of the
light detector determines the resolution.

The position and width of the bands other than the $e/\gamma$-band
can be calculated based on the known quenching factors discussed
above and using the light channel resolutions obtained from the 
fit to the  $e/\gamma$-band. In order to get the width of a
quenched band at a certain energy the light channel resolution
for the actual light energy is used. 

\begin{figure}
  \includegraphics[width=\linewidth]{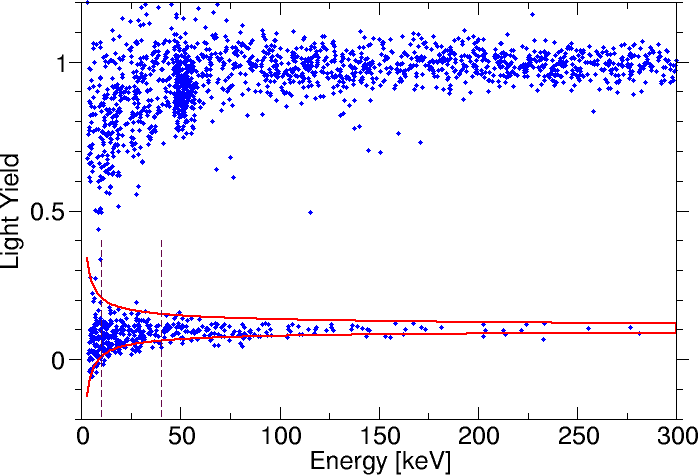}
  \caption{Data obtained with one detector module in a calibration
measurement with an AmBe neutron source, with the source placed
outside the lead shielding. The solid red lines mark
the boundary of the calculated oxygen recoil band 
(\unit[10]{\%} of events are expected above the upper and \unit[10]{\%} below the lower boundary). 
The vertical dashed lines indicate the lower and upper energy
bounds of the WIMP acceptance region as will be introduced in
Section~\ref{sec:latestrun}.}
  \label{fig:neutroncalibration}
\end{figure}

To validate this calculation for quenched bands, we use the data
from a calibration measurement with an AmBe neutron source placed outside the Pb/Cu shielding.
We expect the neutrons to mainly induce oxygen nuclear recoil events.
Fig.~\ref{fig:neutroncalibration} shows the data obtained by one
detector module in this measurement, together with the calculated central \unit[80]{\%} band
for oxygen recoils (\unit[10]{\%} of the events are expected above the upper and
\unit[10]{\%} below the lower boundary).

Nuclear recoil events up to energies of about \unit[300]{keV} 
are observed, with the
spectrum falling off quickly towards high energies. In neutron-nucleus elastic scattering the recoil energy of the nucleus is
inversely proportional to its mass. Thus the highest energy
recoils must be oxygen nuclei. From the ratio of the mass numbers
we then expect the highest energy of calcium recoils  to be around
\unit[100]{keV}. Above
\unit[100]{keV}, we therefore have purely oxygen recoils, and the
 distribution fits well into the calculated
oxygen band. Towards lower
energies, the observed events are still in agreement with the
prediction, although an increasing contribution from calcium
recoils slightly shifts the center of the observed event
distribution to lower light yields.

\section{The Latest Experimental Run} \label{sec:latestrun}

\subsection{Data Set}

The latest run of CRESST took place between June~2009 and
April~2011. It included a neutron test and $\gamma$-cal\-i\-bra\-tions
with $^{57}$Co and $^{232}$Th sources. In total, 18 detector
modules were installed in the cryostat, out of which ten were fully
operated. The remaining modules cannot be employed for a Dark
Matter analysis, principally due to difficulties in cooling the
light detectors.  However, seven additional individual detectors
(six phonon and one light detector) were still operated in order to
tag coincident events (with signals in more than one module).

One of the ten operational modules was equipped with a test
ZnWO$_4$ crystal  and we do not include it in this analysis because
of uncertainties in the quenching factors in this material. Another
operational detector module had unusually poor energy resolution,
with practically no sensitivity in the  WIMP signal region, and was
therefore excluded from the analysis. The data discussed in this
paper were thus collected by eight detector modules, between
July~2009 and March~2011. They correspond to a total net exposure
(after cuts) of \unit[730]{kg\,days}.

\subsection{Observed Event Classes}

\begin{figure}
 \centering
 \includegraphics[width=\linewidth]{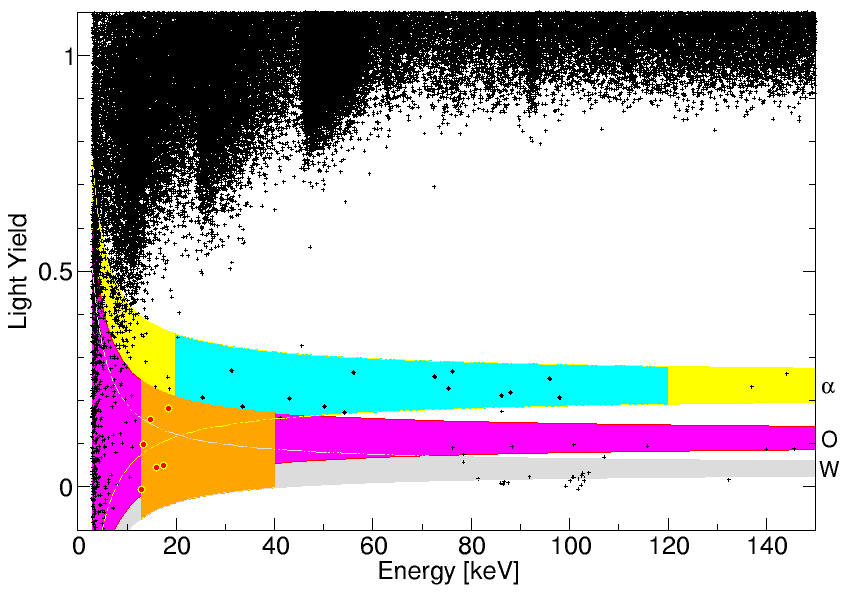}
 \caption{(Color online) The data of one detector module (Ch20), shown in
the light yield vs.\ recoil energy plane. The large number of
events in the band around a light yield of 1 is due to electron and
gamma background events. The shaded areas indicate the bands, where
alpha (yellow), oxygen (violet), and tungsten (gray) recoil events
are expected. Additionally highlighted are the acceptance region
used in this work (orange), the reference region in the
$\alpha$-band (blue), as well as the events observed in these two regions. See text for discussion.}
 \label{fig:yieldplot_example}
\end{figure}

Fig.~\ref{fig:yieldplot_example} shows an example of the data
obtained by one detector module, presented in the light
yield-energy plane. 

 The $e/\gamma$-events are observed around a light yield of 1. The
calculated bands for $\alpha$'s, oxygen recoils, and tungsten
recoils are shown.\footnote{The calcium band is not shown for
clarity. It is located roughly in the middle between the oxygen and
the tungsten bands.} The spread of a band at  each energy
is chosen so that it contains \unit[80]{\%} of the events, that is
 \unit[10]{\%} of the events are expected above the upper boundary
and  \unit[10]{\%} of the events are expected below the lower
boundary. This convention will be used throughout the following
discussion whenever we refer to events being inside or outside of
a band.

Beside the dominant $e/\gamma$-background, we identify several
other classes of events:

Firstly, we observe low energy $\alpha$'s with energies of
\unit[100]{keV} and less. They can be understood as a consequence
of an $\alpha$-contamination in the non-scintillating clamps
holding the crystals.  If the $\alpha$-particle has lost most of
its energy in the clamp before reaching the target crystal, it can
appear at low energy. The rate of such $\alpha$-events differs 
by some factor of two
among the detector modules (see
Section~\ref{sec:qualitativealpha}).

Secondly, Fig.~\ref{fig:yieldplot_example} shows a characteristic
event population in and below the tungsten band around
\unit[100]{keV}. This  is present in all detector modules, albeit
the number of events varies. This population can be attributed to
the lead nuclei from $^{210}$Po $\alpha$-decays on the holding
clamps (see Section~\ref{sec:scintillatinghousing}). The
distribution of
these events exhibits a low-energy tail, with decreasing density
towards lower energies. In spite of this decrease, there are
detector modules (the ones with a high population of such lead events)
in which the tail visibly reaches down to energies as low as a few
tens of keV. 

Finally, low energy events are present in the oxygen, (calcium,)
and tungsten bands at energies up to a few tens of keV, i.e.\ in
the region of interest for the WIMP search. These events will be
the main focus of our discussion in the following. We start by
defining the acceptance region on which the discussion will be
based.

\subsection{Acceptance Region}

Depending on the mass of a  possible WIMP, any of the nuclei in
CaWO$_4$ can be a relevant target for WIMP scattering as
discussed above. We therefore choose our
acceptance region such that it includes all three kinds of nuclear
recoils: it is located between the upper boundary of the oxygen
band and the lower boundary of the tungsten band. This selection
automatically includes the calcium band.

We restrict the accepted recoil energies to below \unit[40]{keV},
since as a result of the incoming WIMP velocities and nuclear form
factors, no significant WIMP signal is expected at higher energies. 
 On the other hand, towards
low energies the finite detector resolution leads to an increasing
leakage of $e/\gamma$-events into the nuclear recoil bands. We
limit this background by imposing a lower energy bound
$E_{\text{acc}}^\text{min}$ in each detector module, chosen such that the
expected $e/\gamma$-leakage into the acceptance region of this
module is one event in the whole data set. Due
to the different resolutions and levels of $e/\gamma$-background in
the crystals, each module is characterized by an individual value
of $E_{\text{acc}}^\text{min}$. Table~\ref{tab:observedevents} lists the values
of  $E_{\text{acc}}^\text{min}$ for all modules.

An example of the resulting acceptance region is shown (orange) in
Fig.~\ref{fig:yieldplot_example} and the events observed therein
are highlighted. In the sum over all eight detector modules, we
then find 67~accepted events, the origin of which we will discuss
in the following. Table~\ref{tab:observedevents} shows the
distribution of these events among the different detector modules. 
{\setlength{\extrarowheight}{2pt}
\begin{table}
\begin{center}
  \begin{tabular}{l|c|c}
     \hline
     module & \unit[$E_{\text{acc}}^\text{min}$ ]{[keV]} & acc.\ events \\ \hline
     Ch05 &   12.3 &  11 \\
     Ch20 &   12.9 &  6 \\
     Ch29 &   12.1 & 17\\
     Ch33 &   15.0 &  6\\
     Ch43 &   15.5 &  9\\
     Ch45 &   16.2 &  4\\
     Ch47 &   19.0 &  5\\
     Ch51 &   10.2 &  9\\ \hline
     \textbf{total}&-&\textbf{67} \\ \hline
  \end{tabular}
\end{center}
\caption{Lower energy limits $E_{\text{acc}}^\text{min}$ of the acceptance regions
and the number of observed events in the acceptance region of each
detector module.}
\label{tab:observedevents}
\end{table}
}
Since $E_{\text{acc}}^\text{min}$  as well as the width of the bands are
module-dependent, different modules have  different sized
acceptance regions and thus different expectations with respect to
background and signal contributions.

\subsection{Backgrounds in the Acceptance Region}

With the above choice of the acceptance region, four sources of
background events can be identified:
\begin{enumerate}
 \item leakage of $e/\gamma$-events at low energies,
 \item $\alpha$-events due to overlap with the $\alpha$-band,
 \item  neutron scatterings which mainly induce oxygen
recoils in the energy range of interest, and
 \item lead recoils from $\alpha$-decays at the surface
of the clamps, degraded to low energy.
\end{enumerate}
In the following, we estimate the contribution of
each of these backgrounds and  finally investigate a possible
excess  above this expectation. When present, such an excess may be the result of WIMP
scatterings in our detectors, or of course an unsuspected
background.

We estimate the backgrounds and the contribution of a possible WIMP
signal in the framework of a maximum likelihood fit that allows a
treatment with all relevant parameters and their uncertainties
simultaneously. However, before introducing this rather abstract
formalism in Section~\ref{sec:likelihoodanalysis}, we give a
qualitative discussion of the backgrounds, with the aim of
clarifying the basic arguments and assumptions.

\section{Qualitative Background Discussion}
\label{sec:qualitativebckdiscussion}

\subsection{\texorpdfstring{$e/\gamma$-Background}{e/gamma-Backgr
ound}}

The lower energy bound of the acceptance region is chosen such that we expect a leakage of
one background $e/\gamma$-event per detector module. These
events are expected to appear mainly at the low energy boundary of
the acceptance region where the overlap between the bands is largest and towards 
the upper boundary, closer to the  $e/\gamma$-band. The
quantitative modeling is discussed in
Section~\ref{sec:likelihoodgamma}.  

\subsection{\texorpdfstring{$\alpha$-}{Alpha }Background}
\label{sec:qualitativealpha}

Since the $\alpha$-band has some overlap with the acceptance
region, some low energy $\alpha$-events may be misidentified as
oxygen or even calcium or tungsten recoils. This will lead to a
certain expectation $n_\text{acc}^{\alpha}$ of background events in the
acceptance region of each module.
 
The energy spectrum $\dd N_\alpha/\dd E$ of the low energy events
in the $\alpha$-band appears to be flat within the available
statistics. In order to estimate $\dd N_\alpha/\dd E$,
we first select a reference region
of the $\alpha$-band which is free of overlap to any other band. 

An example of such a reference region is highlighted (blue)
in Fig.~\ref{fig:yieldplot_example}. To obtain reasonable
statistics, we have chosen the relatively large energy range of
\unit[100]{keV} for the reference region. The low energy limit
$E_\text{ref}^\text{min}$ of the reference region is chosen as low
as possible, while taking into account the increasing
$e/\gamma$-leakage
into the $\alpha$-band towards low energies. Low
$E_\text{ref}^\text{min}$ are desirable, since the reference region
should naturally be close to the acceptance region to minimize
extrapolation errors. On the other hand, influences of the $e/\gamma$-background on the $\alpha$-estimate should be minimized and the value of $E_\text{ref}^\text{min}$ was thus
chosen such that only 0.1 counts of $e/\gamma$-leakage are expected
in the whole reference region for each detector module.
Since the width of the bands varies from module to module, each module has its own value of
$E_\text{ref}^\text{min}$. These values are listed in
Table~\ref{tab:alpha_leakage}.

For our first rough estimate of the alpha background in the
acceptance region, we simply count events in the reference region. We then assume constant $\dd N_\alpha/\dd
E$  and calculate the ratio of $\alpha$-events
expected in the acceptance region to those expected in the
reference region. Scaling the observed number of events in the reference region by this ratio, we arrive at an estimate of the $\alpha$-background in the acceptance region. 

Table~\ref{tab:alpha_leakage} summarizes the observed alpha counts
$n_\text{ref}^{\alpha}$ in the reference region and the resulting
estimates of the alpha background $n_\text{acc}^{\alpha}$ in the
acceptance region of each module. This results in a total expected
$\alpha$-background of about 9.2 events.
 
In the likelihood analysis of Section~\ref{sec:likelihoodanalysis}, we
will relax the assumption of constant $\dd N_\alpha/\dd E$ and also allow for a linear term in the
$\alpha$-energy spectrum. We will, however, see that the simple estimate given here is quite close to the one
obtained with the more sophisticated analysis.

{\setlength{\extrarowheight}{2pt}
\begin{table}
\begin{center}
  \begin{tabular}{l|c|c|c}
     \hline
module &  \unit[$E_\text{ref}^\text{min}$]{[keV]} &
$n_\text{ref}^{\alpha}$& $n_\text{acc}^{\alpha}$ \\ \hline
     Ch05  & 21.7 & 17 &  1.6 \\
     Ch20  & 21.3 & 14 &  1.5\\
     Ch29  & 21.7 & 14 &  1.2\\
     Ch33  & 28.3 &  8 &  0.9\\
     Ch43  & 29.7 &  8 &  0.6\\
     Ch45  & 24.7 &  5 &  0.8\\
     Ch47  & 32.2 &  9 &  1.2\\
     Ch51  & 18.3 & 18 &  1.4\\ \hline
        \textbf{total}& - &\textbf{93} & \textbf{9.2} \\ \hline
  \end{tabular}
\end{center}
\caption{Lower energy limits $E_\text{ref}^\text{min}$ of the
$\alpha$-reference regions, observed alpha counts 
$n_\text{ref}^{\alpha}$ in the reference regions, and the resulting
(rough) estimates of the alpha background  $n_\text{acc}^{\alpha}$
in the acceptance regions of all detector modules.}
\label{tab:alpha_leakage}
\end{table}
}

\subsection{Neutron Background}

\subsubsection{Introduction}
Throughout our discussion, we distinguish two different classes of
neutron production mechanisms: 

Firstly, free neutrons can be emitted by radioactive processes, in
particular spontaneous fission (s.f.) of heavy elements or
$(\alpha,n)$ reactions on light nuclei. Such neutrons typically
have energies up to a few MeV, for which the polyethylene shielding
provides a very efficient moderator. Monte Carlo simulations
suggest that neutrons from  s.f. and $(\alpha,n)$ reactions in the
rock outside the experiment thus only constitute a negligible background at
the level of $10^{-5}$ events per
kg\,d~\cite{Wulandari2004_neutronflux}. Therefore, such neutrons
are a possibly relevant background only if they are emitted \emph{inside}
the neutron shielding, e.g.\ by s.f. of $^{238}$U in the lead
shielding, or
by $(\alpha,n)$ reactions or s.f. in the copper shielding. We
consider such neutrons here, even though Monte Carlo simulations
predict only a negligible background in the acceptance region due
to these processes, at a level of $10^{-3}$ events per
kg\,d~\cite{Scholl2011_phd}.  

Secondly, neutrons can also be produced by  muons,
either in the lead or copper shielding  or in the
surrounding rock. In the former case, the muon will mostly be
tagged by the muon veto enclosing the Pb/Cu shielding. However,
there is
a small probability that the muon is missed by the veto because of
the hole on top of the setup,  as mentioned above (cf.\ Fig.~\ref{fig:cresstsetup}). Such a muon may create a shower
of neutrons inside the PE shielding which  then reach the
detectors. On the other hand, muon-induced neutrons created
\emph{outside} the neutron shielding may penetrate the polyethylene
layers if they are energetic enough. Such high energy neutrons then
have a  high probability to scatter inelastically in the Pb/Cu
shielding and to create  secondary neutrons and gammas.
Ultimately, this leads to events with similar characteristics as
when the shower is directly induced by a muon \emph{inside} the Pb/Cu
shielding.

\subsubsection{Experimental Information}

It is a characteristic feature of neutrons that they can lead to coincident events in more than one detector module at the same time, with at least one module registering a nuclear recoil.  A given neutron source will thereby lead to events with a characteristic ratio between single and coincident scatterings. If this ratio is known, the
observed coincidences can be exploited to estimate the expected
number of single neutron (background) events. This is the basic concept pursued in the following. 

We shall base our discussion on
two sources of information. One is the results of the neutron test
with an AmBe neutron source already mentioned above. The second is the examination of
events in coincidence with incoming muons, i.e.\ muon veto triggers
accompanied with a signal in the acceptance region of a module.
These two ``calibration measurements'' are used to infer the properties of neutron backgrounds
originating from ambient neutrons and from  muons escaping the veto
and interacting in the apparatus. We emphasize that Monte Carlo or other external
information is helpful, but does not enter into our quantitative
estimates.

\begin{figure}
\centering
\includegraphics
[width=0.9\linewidth]{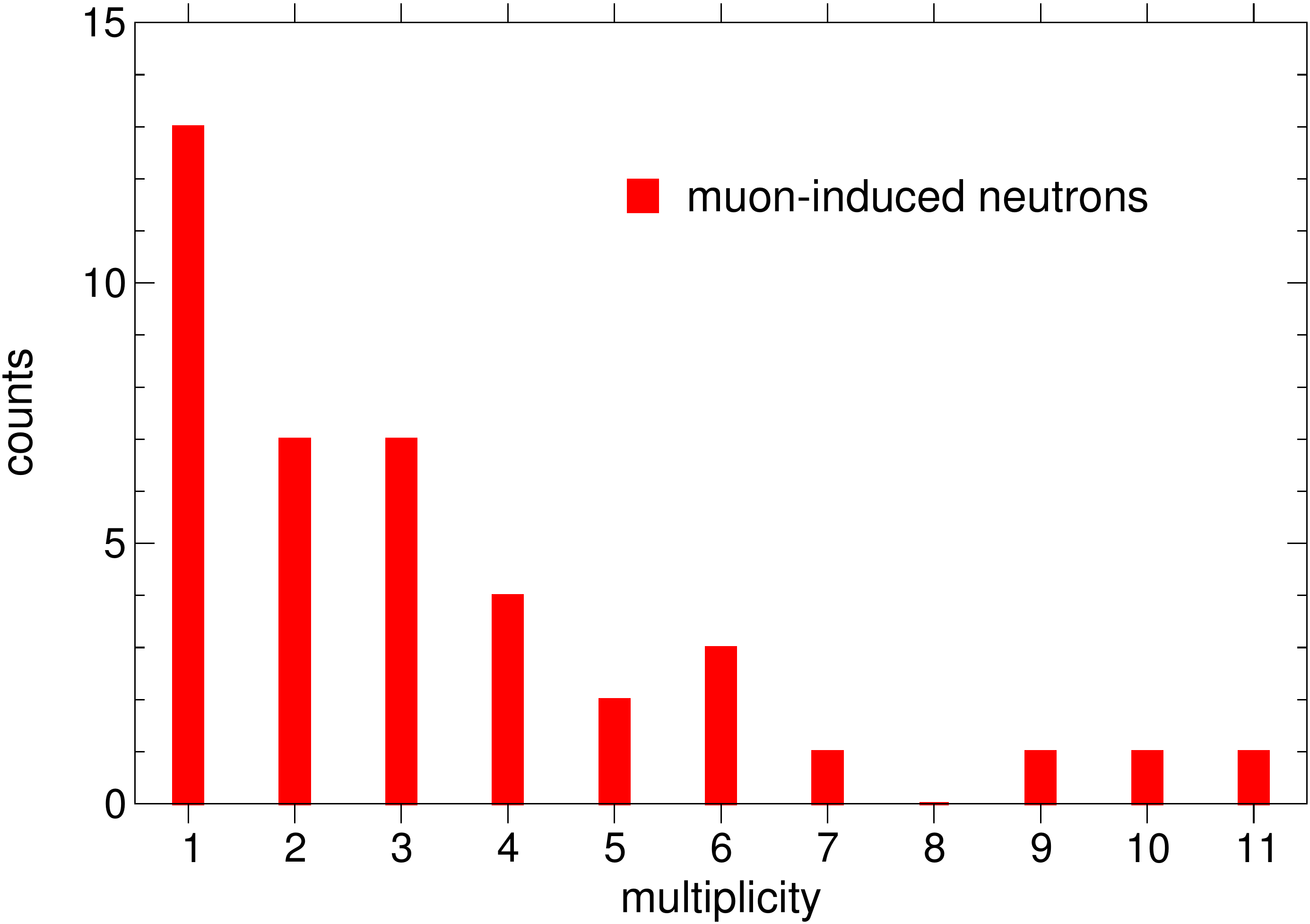}
\vspace*{0.5cm}

\includegraphics[width=0.9\linewidth]
{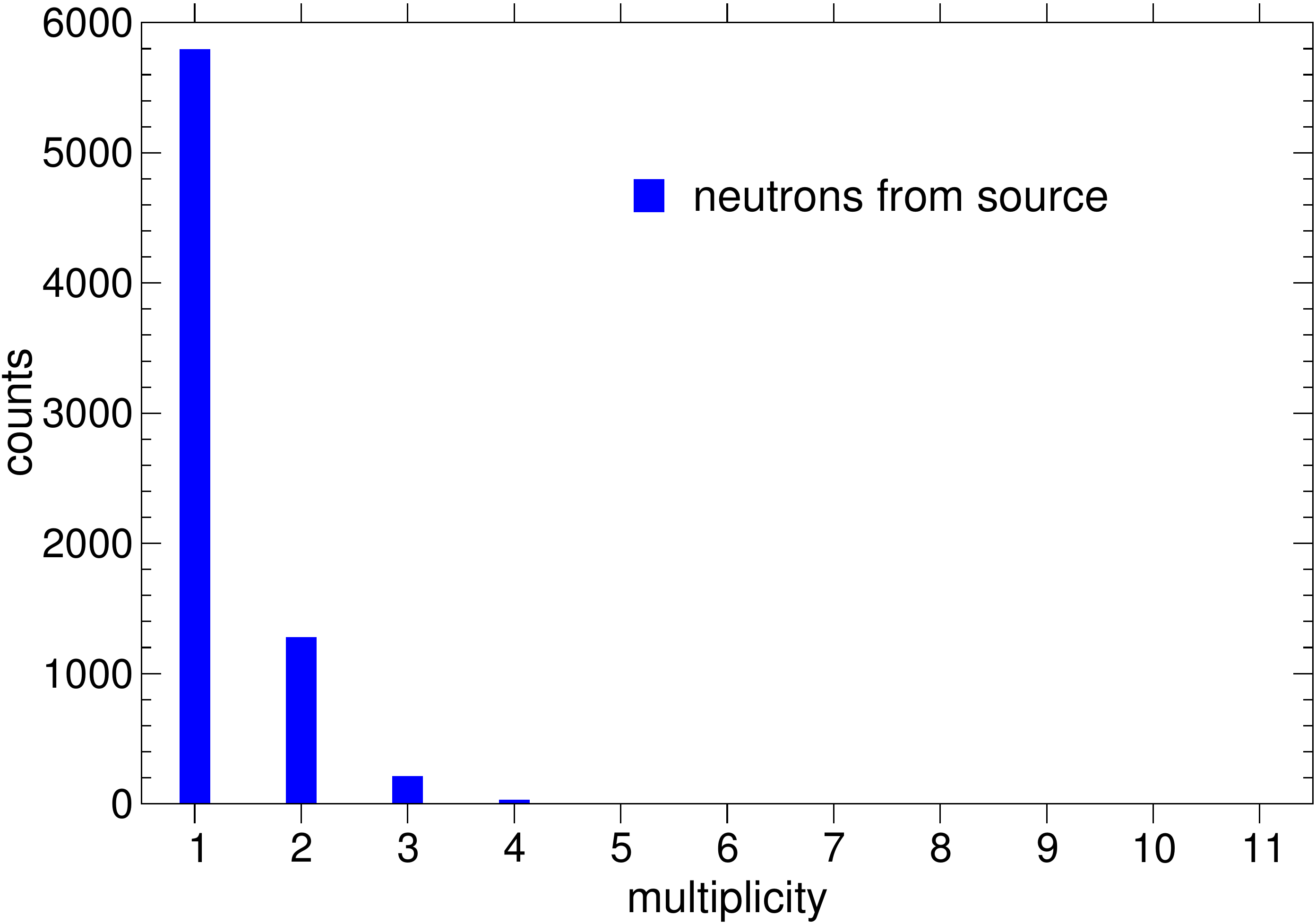}
\caption{Multiplicity (i.e.\ the number of modules which
  triggered in coincidence) of events which include a nuclear
  recoil in the acceptance region of at least one 
  detector module. Top: muon-induced events, bottom: events from an
AmBe neutron source.}
\label{fig:multiplicities}
\end{figure}

For each type of neutron mechanism, we use the corresponding calibration to infer the
typical structure of the events with respect to their \emph{multiplicity}, defined as the number of detector modules triggering at the same time (within a time window of \unit[5]{ms}). 
For muon-induced neutrons, we
observe 40 events coincident with a muon veto trigger and with at least one detector module having
a signal in its acceptance region. The multiplicities of
these events are shown in the top histogram of
Fig.~\ref{fig:multiplicities}. On the other hand, the multiplicities of the events induced by the AmBe neutron source (placed at various positions outside the Pb/Cu shielding but inside the neutron shielding) are given in the bottom histogram of
Fig.~\ref{fig:multiplicities}.  

Events due to muon-induced neutrons  show an obviously
 higher average multiplicity than events from the  neutron test
source. This evidently results from  muon-induced cascades, leading
to  neutrons and gammas  in different detector modules at the same time. 

With this information at hand, we turn back to the Dark Matter data set. In addition to the 67 accepted events, these data contain three events in which several detector modules triggered in coincidence, 
with at least one module registering an event in its acceptance
region. Two of these events have a
\emph{multiplicity} of three (i.e.\ three modules triggered), while
in the third event five modules triggered in coincidence. Such
coincidences which include at least one nuclear recoil must involve
a  neutron, and perhaps $\gamma$'s. Accidental coincidences may be
neglected in view of the low overall counting rate.

Based on these three coincident events we can use the calibration information given above to scale up to the number of expected \emph{single} scatters for each type of source. 

For a first rough estimate of the level of neutron background, we
start with the three observed coincidences and neglect 
their precise multiplicities. From the histograms in
Fig.~\ref{fig:multiplicities}, we find that muon-induced neutrons
are characterized by a ratio of single to coincident scatterings of
about 0.5, while the  neutron source test gives a ratio of
about 3.8. Consequently, if we assume that only muon-induced
neutrons are present in the experiment, this would lead to an
estimated number of single scatterings of $3 \cdot 0.5 = 1.5$. On
the other hand, in case the neutron background purely comes from a
radioactive source, the same estimate gives a background
expectation of about  $3 \cdot 3.8 = 11.4$ single events.

In reality, both types of neutrons may be present, and the above
limiting cases show the extremes between which the expected neutron
background can lie according to this estimate. Our likelihood
analysis of Section~\ref{sec:likelihoodanalysis} takes into account
the more detailed information of the individual event
multiplicities in order to clarify the contributions of the two
types of neutron  sources to the total background. We will,
however, see that the result is compatible with the simple
estimates  of the limiting cases given here.

An independent aspect of the neutron background concerns the
corresponding recoil energy spectrum. Within our narrow accepted
energy range, the energy spectra induced by the two types of neutron events
are found to be very similar, according to the calibration data discussed above. The spectrum can be parametrized by a simple
exponential $\dd N_\text{n}/\dd E \propto \exp{(-
E/E_\text{dec})}$. We
determine the parameter $E_\text{dec}$ from a fit to the spectrum
obtained in the AmBe neutron calibration run. In the
energy range between \unit[12]{keV} to \unit[40]{keV} we obtain
\mbox{\unit[$E_\text{dec}=(23.54 \pm 0.92)$]{keV}}. 

This similarity in the spectra induced by neutrons  from the two quite
different sources (in agreement with Monte Carlo results
\cite{Scholl2011_phd}) 
indicates how the Pb/Cu shielding surrounding the detectors will moderate
an incoming neutron flux regardless of its origin. The primary spectrum of the neutrons is washed out by inelastic scatterings in the shielding. This finding supports
our use of the results of the neutron calibration to estimate the
effects of a  general neutron background. The only exception to
this argument might be a neutron-producing contamination in
close vicinity of the detectors. In this case, we would expect a 
recoil spectrum reaching to much higher energies and
fewer singles for a given number of coincidences. In this case, the application of our above calibration results would lead to a conservative neutron background estimate.

\subsection{Lead Recoil Background}
\label{sec:leadrecoils_qualitative}

\begin{figure}
\centering
 \includegraphics[width=\linewidth]{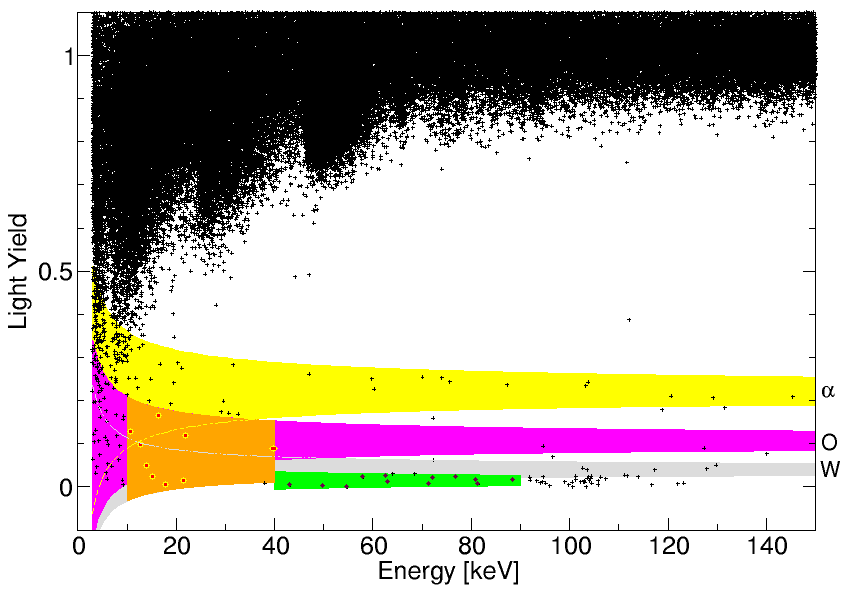}
 \caption{(Color online) The data of detector module Ch51, shown in
the light yield vs.\ recoil energy plane. Again, the shaded areas
indicate the bands, where alpha (yellow), oxygen (violet), and
tungsten (gray) recoil events are expected. Additionally
highlighted are the acceptance region (orange), the region  where
lead recoils with energies between 40 and \unit[90]{keV} are
expected (green), and the events observed in these regions. The highlighted lead recoil region (green) serves as a reference region for
estimating the $^{206}$Pb recoil background.}
 \label{fig:ch51_yield_plot}
\end{figure}

To illustrate the lead recoil background from $^{210}$Po decay,
Fig.~\ref{fig:ch51_yield_plot} displays the data set of a different
detector module as in Fig.~\ref{fig:yieldplot_example}. Compared to
Fig.~\ref{fig:yieldplot_example}, a
more prominent population of $^{206}$Pb recoils below the tungsten
band is visible, with a rather long  tail extending down to the
acceptance region. Since the lead band and the
acceptance region overlap considerably, a leakage of some
$^{206}$Pb events into the acceptance region cannot be excluded.

For an estimate of this background, we follow a similar strategy as
for the $\alpha$-background. We define a reference
region for each detector module which contains predominantly
$^{206}$Pb recoils, and model the spectral energy density $\dd
N_\text{Pb}/\dd E$ in this region. This model is then extrapolated
into the energy range of the acceptance region. 

As a reference region, we choose the lead recoil band at energies
above the acceptance region, where a possible WIMP signal cannot
contribute. In some detector modules with wider bands, the lead
band still overlaps with the oxygen band around the lower edge of
this energy range. In this case, we additionally restrict the
reference region to the lower part of the lead band without overlap
with the oxygen band in order to be independent of possible
neutron-induced events on oxygen. The event distribution
of the Pb recoils peaks at the full lead recoil energy of \unit[103]{keV} and  the
upper boundary of the
reference region is set at \unit[90]{keV}
so that it covers the low energy tail. An
example of the resulting reference region is  highlighted
in green in Fig.~\ref{fig:ch51_yield_plot}.
Table~\ref{tab:Pb_leakage} summarizes the counts
$n_\text{ref}^\text{Pb}$ observed in the reference region of each
detector module.

{
\setlength{\extrarowheight}{2pt}
\begin{table}
\begin{center}
  \begin{tabular}{l|r}
     \hline
module &  $n_\text{ref}^{\text{Pb}}$ \\ \hline
     Ch05  & 17 \\
     Ch20  &  6 \\
     Ch29  & 14 \\
     Ch33  &  6 \\
     Ch43  & 12 \\
     Ch45  & 15 \\
     Ch47  &  7 \\
     Ch51  & 12 \\ \hline
\textbf{total}& \textbf{89} \\ \hline
  \end{tabular}
\end{center}
\caption{Observed counts $n_\text{ref}^{\text{Pb}}$ in the lead
reference regions of the detector modules.}
\label{tab:Pb_leakage}
\end{table}
}

\begin{figure}
 \centering
 \includegraphics[width=\linewidth]{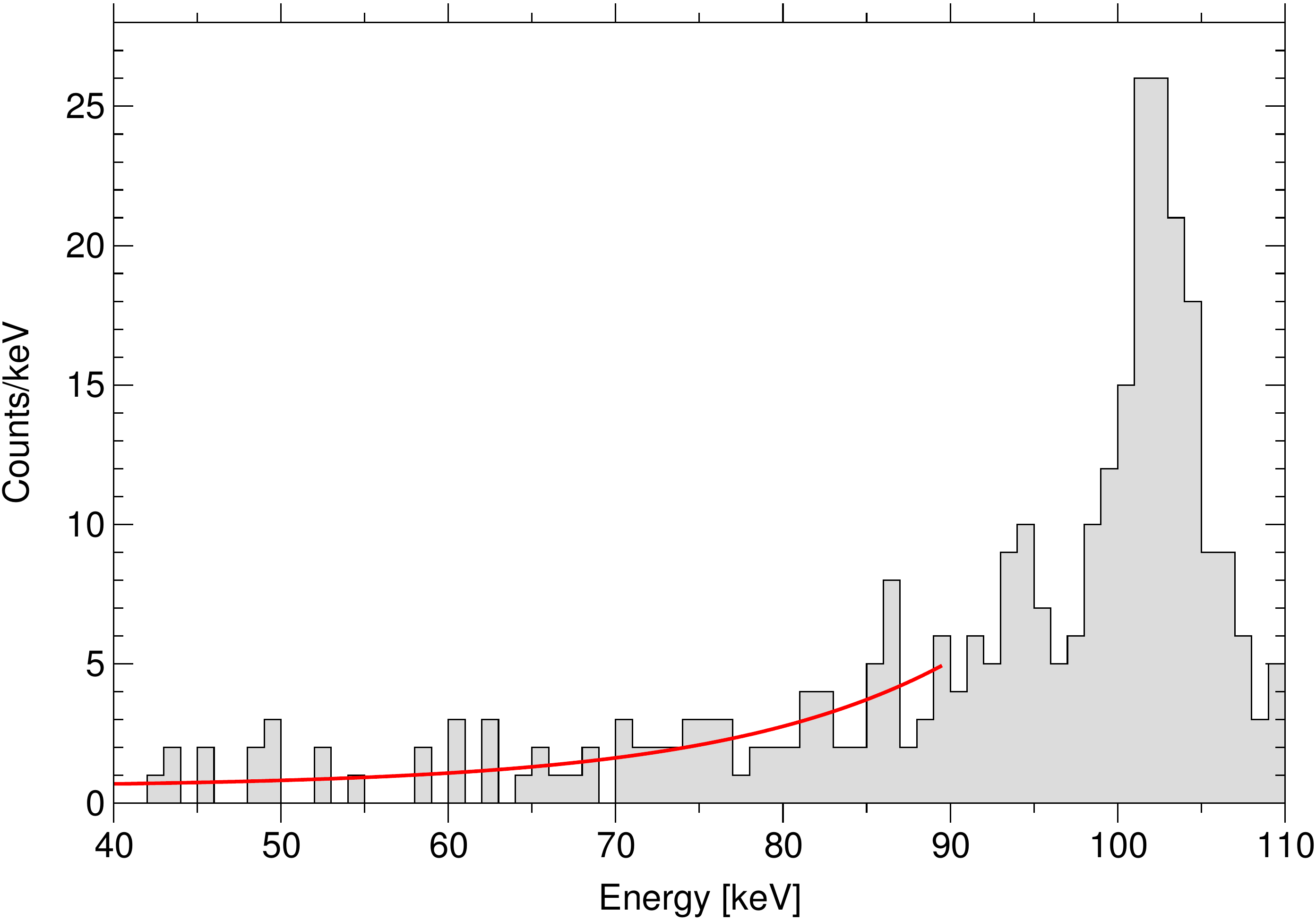}
 \caption{Energy spectrum of events in the $^{206}$Pb reference
region of all detector modules. Events occurring in the lead band
at energies above the upper energy limit of the reference region
are included to show the peak at the nominal recoil energy of 103
keV. The curve (red) is a fit of the histogram with an exponential
plus a constant term, in the energy range of the reference region.}
 \label{fig:Pb_histogram}
\end{figure}

Fig.~\ref{fig:Pb_histogram} presents the energy spectrum of the
events found in the $^{206}$Pb reference regions of all detector
modules, but includes also lead recoils with higher energies to
illustrate the peak at the full nominal recoil energy of
\unit[103]{keV}. In the energy range of the reference region (below
\unit[90]{keV}), the tail of the distribution can be modeled by an
exponential decay on top of a constant contribution:

\begin{equation} \label{eq:PbFit}
   \frac{\dd N_\text{Pb}}{\dd E}(E) = A_\text{Pb} \cdot \left[
C_\text{Pb}+\exp\left(\frac{E-\unit[90]{keV}}
{E^\text{Pb}_\text{decay}} \right) \right].
\end{equation}
 
For a first rough estimate of the recoil background, we simply fit
such a  function to the spectrum of
Fig.~\ref{fig:Pb_histogram}. The red line shows the result of this
fit with the parameters 
$A_\text{Pb}=\unit[4.53]{counts/keV}$, $C_\text{Pb}=0.13$ and
$E_\text{decay}^\text{Pb}=\unit[13.72]{keV}$. This model then needs
to be extrapolated into the energy range of the acceptance region.

\begin{figure}
 \centering
 \includegraphics[width=\linewidth]{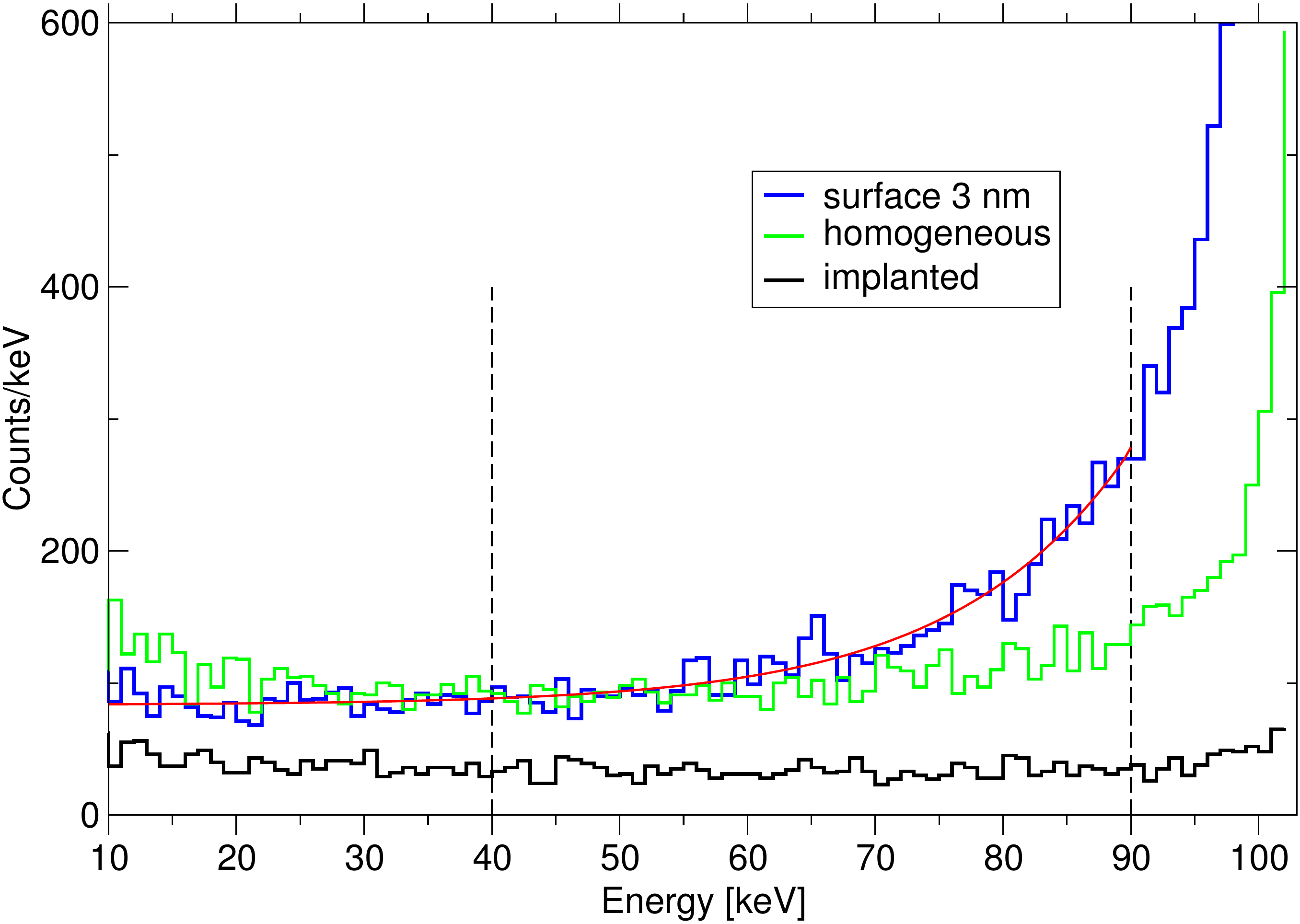}
 \caption{Energy spectrum of $^{206}$Pb recoils calculated with the
SRIM package for three different depth profiles of the alpha
emitting $^{210}$Po parent nuclei:  (top) distribution peaking at
the surface with a \unit[3]{nm} exponential decay length, (middle) uniform
distribution throughout the volume, and (bottom) a depth profile
resulting from adsorption of $^{222}$Rn at the surface of the
clamps, with a subsequent implantation by the two alpha decays
which follow in the $^{238}$U decay chain. The full (red) curve is the result of a fit of Eq.~\ref{eq:PbFit} to the top spectrum, in the
energy range of the reference region from 40 to \unit[90]{keV}. The extrapolation of this fit into the acceptance region below
\unit[40]{keV} is shown.
}
 \label{fig:SRIM_spectra}
\end{figure}

To check the validity of this extrapolation we used the SRIM
package \cite{SRIM_software} to simulate the  energy spectra
expected for three different depth distributions of the 
$^{210}$Po mother nucleus. These distributions were: an
exponential profile with \unit[3]{nm} decay length peaking at the surface,
a uniform distribution in the volume, and finally the depth
distribution resulting from  implantation due to preceding alpha
decays. For the latter case the implantation profile was also
calculated with SRIM, assuming that $^{222}$Rn is first adsorbed on
the surface of the clamps holding the crystals, followed by two subsequent alpha decays to $^{210}$Po.
The results of the three simulations are shown in
Fig.~\ref{fig:SRIM_spectra}. 

An important result of the simulation is that none of the
calculated spectra of the Pb recoils rises significantly towards
low energies within the range of our acceptance regions. The
simulated spectra for the uniform as well as the implantation
profile are rather flat between 40 and \unit[90]{keV} compared to
the data in Fig.~\ref{fig:Pb_histogram}. However, the energy
spectrum from the distribution peaking at the surface has a tail
similar to that observed in the data. The curve in
Fig.~\ref{fig:SRIM_spectra} is the result of a fit of
Eq.~\ref{eq:PbFit} to this spectrum in the energy range of the
reference region between 40 and \unit[90]{keV}. Its exponential
decay of $E_\text{decay}^\text{Pb}=\unit[13.6]{keV}$ agrees within
errors with the value obtained from the fit of the data in
Fig.~\ref{fig:Pb_histogram}. Below \unit[40]{keV} the curve in
Fig.~\ref{fig:SRIM_spectra} shows the extrapolation of this function into the accepted energy range. The very good
agreement of extrapolation and  simulated data in the energy range
of the acceptance region justifies the use of this type of
extrapolation for our estimate of the Pb background.

For a rough first result, we take a typical energy range for the
acceptance region of 12~to~\unit[40]{keV} and estimate the number
of Pb recoils in it. From the extrapolation
of the fit function in Fig.~\ref{fig:Pb_histogram} we calculate
about 17~events of $^{206}$Pb background in the acceptance region.
Of course, this simple estimate neglects small differences in the
overlap of the lead band with the acceptance region in different
detector modules, as well as the acceptance reduction in the
reference region due to partial overlap of lead and oxygen bands.
Nevertheless, the result is very close to the final value that we
will obtain from the full likelihood analysis, which performs the
background estimate module-wise and hence takes such differences
into account.

\section{Maximum Likelihood Analysis}
\label{sec:likelihoodanalysis}

In the previous section, the qualitative principles of our
background estimates were discussed. This section explains how
we formulate these concepts quantitatively. We choose the framework
of a likelihood analysis to estimate the unknown parameters of our
backgrounds and as well as the ones of a possible signal. This
formalism also allows us to take into account and propagate the corresponding
uncertainties. We first give a general overview of the formalism
before focusing in detail on the treatment of the different
backgrounds in the current measurement. 

\subsection{General Concepts} \label{sec:likelihoodconcepts}

The maximum likelihood fit is based on a  parameterized
model of the backgrounds and a possible signal. The parameters are then varied
to find the values for which the model most likely reproduces the data. 

As we will see, this procedure allows one to take
into account the actual position of the events in the light yield-energy
plane, and not  merely their presence in a broad acceptance region.
Also, measurements from reference regions, outside
the acceptance regions, may be introduced to help determine some of
the parameters. Furthermore one may, in the framework of the method, give a
quantitative estimate of how well a best set of parameters is determined.

The maximum likelihood method itself however gives no direct indication of the quality of the resulting fit. To deal with this question, we shall attempt to give a characterization of the quality-of-fit using the so-called $p$-value.
 
\subsubsection{Likelihood Function}
The basic input to the analysis is a model for each source $X$ that
may contribute to our accepted events, be it background or a
possible signal. Such a model describes the \emph{expected}
distribution of events in the $(E,y)$ plane (where $E$ denotes the
recoil energy and $y$ the light yield). 

We formulate such a model in terms of  a density $\rho_X(E,y\, |\,
\mathbf{p}_X )$ in the $(E,y)$ plane. The  number of events
expected
from source $X$ in any subspace of the $(E,y)$ plane is then simply
given by the integral of $\rho_X$ over this region.
 $\rho_X$ depends on a set of unknown parameters
summarized in the vector $\mathbf{p}_X$. These are the parameters
to be varied. 

In our case, each such density function is the product of two
components: the expected recoil energy spectrum $\dd N_X/\dd E$ 
and a function that describes the expected event
distribution in the light yield coordinate at each recoil energy
$E$.

As discussed above, we use 
a Gaussian distribution  for the light yield, where the
center of the distribution is given by the respective quenching
factor  and the width is essentially determined by the
resolution of the light channel. 

 Since different detector modules have different
resolutions, the densities $\rho_X$ need to be defined for each
module individually. We therefore add an additional
 index $d$  for "detector module" to each density.
Assuming that we have considered all significant sources of events,
the total density for  module
$d$ is
\begin{equation}
 \rho^d(E,y\,|\,\mathbf{p}) = \rho^d_\gamma + \rho^d_\alpha +
\rho^d_\text{neutron} + \rho_\text{Pb}^d + \rho^d_\chi,
\end{equation}
where we have taken into account the four backgrounds discussed
above, plus a possible WIMP signal "$\chi$". The vector
$\mathbf{p}$
summarizes all the unknown parameters. 

 We now have \emph{observed} a set of
events in the acceptance region of each module, located at the
positions $(E_i,y_i)$. The functions $\rho^d$ can be fitted to the
event distribution in the respective module and, from this the
most likely values of the unknown parameters can, in principle, be
determined.
 The normalizations of the $\rho$ depend on the variable
parameters and give the total number of expected events
$\mathcal{N}$ via 
\begin{equation} \label{eq:extendedlikelihoodnormalization}
  \mathcal{N}(\mathbf{p}) := \sum_d \left( \iint_{\text{acc.\
region}} \rho^d(E,y\,|\,\mathbf{p}) \ \dd E \,\dd y \right), 
\end{equation}
where the integral runs over the acceptance region of module $d$
and thus yields the total expected number of accepted events in
this module.

The $\rho^d$ are thus no probabilty densities, but they can be used in the so-called \emph{extended maximum
likelihood} formalism \cite{Barlow1990_ExtendedML} to formulate the likelihood function which then takes the form
 
\begin{equation} \label{eq:extendedlikelihood}
 \mathcal{L}_\text{acc}(\mathbf{p}) = \left[ \prod\limits _d
\prod\limits _i \rho^d(E_i,y_i \,|\, \mathbf{p} ) \right] \cdot
\exp[-\mathcal{N}(\mathbf{p} )],
\end{equation}
with the outer product running over the detector modules and the
inner one over all accepted events in the respective module. The
exponential factor takes into account the normalization of the
$\rho^d$.

We emphasize that, in (\ref{eq:extendedlikelihood}), we directly
evaluate $\rho^d$ at the position of each observed event and no
binning is involved. By fitting the density functions to the
observed event coordinates, one makes use of the full available
information from the two-channel measurement.  Using the
experimental data, the fit then finds  the most likely values of
the parameters, including those for the backgrounds and  a
possible signal.

\subsubsection{Reference Regions}
In practice, some of the unknown parameters in $\mathbf{p}$ are not
sufficiently constrained by the observation in the acceptance
region alone. In such cases, additional observations (``reference
measurements``) can be exploited which are made either in other
regions of the measurement parameter space (for example in the
$\alpha$-reference region introduced in the previous section), or
in completely different experiments or measurement channels. Each
such reference measurement yields a separate likelihood function
which constrains a subset $\mathbf{p'}$ of $\mathbf{p}$, and,
assuming all measurements are independent, the total likelihood is
simply given by the product
\begin{equation}
  \mathcal{L}_\text{tot}(\mathbf{p}) =
\mathcal{L}_\text{acc}(\mathbf{p}) \cdot
\underbrace{\mathcal{L}_\text{ref1}(\mathbf{p'_1}) \cdot
\mathcal{L}_\text{ref2}(\mathbf{p'_2}) \cdot
\ldots}_{\text{reference measurements}} \: .
\end{equation}

This concept can also be exploited in order to take into account
the effects of uncertainties on otherwise fixed quantities, like
e.g.\ the quenching factors of O, Ca and W, which enter the analysis
and have been measured previously.  To this end, these quantities
are added to $\mathbf{p}$ as free parameters and, at the same time,
an additional likelihood term is included for each of them which
models the uncertainty of the quantity in question (e.g.\ a
Gaussian around the best estimate, with the width given by the
error on this estimate).

Finally, maximizing the total likelihood function
$\mathcal{L}_\text{tot}(\mathbf{p})$ leads to estimates for all
unknown parameters $\mathbf{p}$. Most of the time, however, one is
only interested in a subset of these values, the others just being
nuisance parameters required to construct the likelihood. In our
case, the only parameters of interest are the WIMP-nucleon cross
section $\sigma_\text{WN}$ and possibly (if we find a clear signal)
the WIMP mass $m_\chi$. Our aim is to derive confidence intervals
for these relevant quantities, taking into account the nuisance
parameters and their uncertainties.

\subsubsection{WIMP Parameters}
In a first step, we will thereby only concentrate on
$\sigma_\text{WN}$ and ask whether our measurement indicates
$\sigma_\text{WN}>0$  significantly. All other
parameters shall be summarized in the vector $\mathbf{p'}$. This
question corresponds to a test of the null-hypothesis
$\sigma_\text{WN}=0$ with the model. 

As a convenient test statistic, we employ the \emph{likelihood
ratio}
\begin{equation}
\Lambda(\sigma_\text{WN}=0) :=
\frac{\mathcal{L}_\text{tot}(\sigma_\text{WN}=0,
\widetilde{\mathbf{p'}})}{\mathcal{L}_\text{tot}(\hat
\sigma_\text{WN}, \widehat{\mathbf{p'}})}.
\end{equation}
Here, $\hat \sigma_\text{WN}$ and $\widehat{\mathbf{p'}}$ in the
denominator are the maximum likelihood estimators of
$\sigma_\text{WN}$ and $\mathbf{p'}$, respectively, and
$\widetilde{\mathbf{p'}}$ is the conditional maximum likelihood
estimator under the condition $\sigma_\text{WN}=0$. $\Lambda$ is
hence a measure of how ''signal-like`` our observation is, in the
sense that small values indicate a considerably better description
of the data \emph{with} an allowed  WIMP contribution than without.

According to Wilk's theorem \cite{Wilks1938_chi2}, if
$\sigma_\text{WN}=0$ were the true assumption, then the quantity 
\begin{equation}
q:= -2 \ln \Lambda(\sigma_\text{WN}=0)
\end{equation}
would have a $\chi^2$-probability distribution for one degree of
freedom in the limit of large statistics. We have verified with a
Monte Carlo simulation that this approximation holds well for the
case of our analysis. This allows for a simple calculation of the
statistical significance $S$, with which we can reject the null
hypothesis $\sigma_\text{WN}=0$ when having observed a certain
value $q_\text{obs}$:
\begin{equation}
 S = \sqrt{q_\text{obs}}.
\end{equation}
Higher values of $S$ obviously imply smaller probabilities to get
data as "signal-like'' as the observed ones, although there is no
true signal present.

In a second step, if $\sigma_\text{WN}>0$ can be established with
sufficient significance, the above approach can be generalized and
$m_\chi$ can be treated as a second parameter of interest. The aim
is then to calculate a confidence region in the $(m_\chi,
\sigma_\text{WN})$ plane.
Such confidence regions are given by the contours on which the
likelihood has decreased by a certain factor $\delta \mathcal{L}$
with respect to the maximum, provided that all parameters other
than $m_\chi$ and $\sigma_\text{WN}$ are refitted in each point.
The value of $\delta \mathcal{L}$ that yields the desired
confidence level can thereby be obtained from a
$\chi^2$-distribution for two degrees of freedom. In fact, the same
procedure can be applied to calculate confidence regions for any
combination of $n\geq1$ parameters, using the $\chi^2$-distribution
for $n$ degrees of freedom. The one-dimensional case is commonly
known from its implementation in the MINOS program of the MINUIT
software package~\cite{james1975_minuit}.

\subsubsection{$p$-Value}
As mentioned above, the significance resulting from the
above likelihood ratio test is, in itself, not a good measure of
the quality of the  fit to the data. High significances do not
automatically imply a good agreement between the fitted model and
the data, so that the quality-of-fit has to be determined in an
independent step. One possibility of doing so is described in
\cite{Beaujean2010_pvalues} and we adopt an analogous procedure
here: 

The $(E,y)$-plane (or a subspace of it) is divided into
two-dimensional bins, separately for each detector module. In each
bin $i$, the expected number $e_i$ of events can be calculated by
integrating the density $\rho^d$. To go with this expectation, we
have certain numbers of \emph{observed} events in the bins. We can
then determine the total probability of this observation by
assuming an independent Poisson process with expectation $e_i$ in
each bin $i$ and multiplying all corresponding Poisson
probabilities.

Let our real observation have such a probability $P_\text{obs}$.
Independently, we can then generate artificial data sets where, in
each bin $i$, the generated number of events is Poisson distributed
with the expectation $e_i$. We employ a Markov Chain Monte Carlo
for a fast generation of these data sets. Each generated set $s$
can then again be characterized by a certain probability $P_s$.
From the distribution of the $P_s$, we finally determine the
probability to get a data set with probability less or equal to
$P_\text{obs}$. This is called the $p$-value and can be used to 
describe the quality-of-fit.

Often, as in our case, the model used to generate the artificial
data sets is the result of a previous fit to the real observed
data, where parameter values have been tuned for an optimal
description of this observation. The $p$-value as obtained above
needs to be corrected for this bias and a procedure of doing so is
suggested in \cite{Beaujean2010_pvalues}. We perform this
correction, but we note that its influence is naturally small when
the number of free fit parameters is small compared to the number
of bins considered.

\vspace{1cm}
Having now established the formal framework of our analysis, we
will focus in the following on the concrete  construction of the
densities as well as the required reference measurements for each
considered source of events.

\subsection{\texorpdfstring{$e/\gamma$}{e/gamma}-Background}
\label{sec:likelihoodgamma}

The distribution of the $e/\gamma$-events in the $(E,y)$ plane is
observable directly in the Dark Matter data set. Due to the large
number of such events, statistical fluctuations are negligible and
we use the observed distribution directly to fix the densities
$\rho^d_\gamma(E,y)$. The fit of the energy-dependent Gaussian
width of the $e/\gamma$-band as outlined above thereby gives the
parametrization of the light yield distribution of the events. On
the other hand, we use a histogram of the observed recoil energies
(bin width \unit[0.1]{keV}) as the expected energy spectrum and
interpolate it to arrive at a continuous function. In the energy
range of interest, the statistical uncertainties of the bin
contents of this histograms are below $10^{-2}$. The resulting
densities $\rho^d_\gamma$ can therefore be treated as fixed
contributions to the total density in each module, with no free fit
parameters.    

\subsection{\texorpdfstring{$\alpha$-}{Alpha }Background}

As discussed above, the observation of events in overlap-free
regions of the $\alpha$-band indicates an approximately flat energy
spectrum of the $\alpha$-background in our data. Nevertheless, to
account for a possible small systematic energy dependence, we model
the energy spectrum in each detector module $d$ as
\begin{equation}
 \frac{\dd N_\alpha^d}{\dd E} (E) =  c_\alpha^d + m_\alpha \cdot E,
\end{equation}
introducing the fit parameters $c_\alpha^d$ and $m_\alpha$.
$c_\alpha^d$ is module-specific and describes the level of $\alpha$-activity observed in the respective module, while we assume the
possible slope $m_\alpha$ to be common to all modules. It reflects
the properties of the alpha energy loss before interaction in the
target.

Obviously, the above fit parameters are only weakly constrained by
the events observed in the acceptance region, where other sources
contribute as well. Therefore, we additionally consider the
$\alpha$-reference region for each detector module as introduced in
Section~\ref{sec:qualitativebckdiscussion}. It is this region that
mainly influences the estimates of the $\alpha$-parameters.
Technically, we simply extend the inner product in
Eq.~(\ref{eq:extendedlikelihood}) to also run over the events
observed in the reference region and modify the integral in
Eq.~(\ref{eq:extendedlikelihoodnormalization}) accordingly. Since
the contributions from other sources are small in the reference
regions, this modification mainly leads to additional constraints
on the $\alpha$-parameters. 

A second benefit of using these reference regions is that the
low-energy quenching factor for $\alpha$-particles can be left free
in the fit. This is particularly important, since our observation
of alphas in this data set is the best measurement available for
this quantity. The quenching factor is  automatically constrained
by the events observed in the reference region and the
uncertainties of this measurement are directly built into the
likelihood. Of course, in order to define the reference region, a
reasonable starting assumption for the quenching factor is
required. We obtain it simply from the one-dimensional light yield
distribution of the low-energy $\alpha$-events. If this starting
assumption is not too far from the true value, the likelihood fit
can give a good estimate for the quenching factor, even though it
is based on the events previously selected under the assumption of
a slightly different starting value. 
We find that the result is indeed very robust against reasonable
changes in the starting value. 

We also confirmed that the result is very robust against some
reasonable changes of the energy range of the reference regions. A
selection of the same energy range for all detector modules of 35
to \unit[135]{keV}, or selecting a 150 keV wide reference region
with the lower energy limits of Table~\ref{tab:alpha_leakage} did
not change results in any relevant way.

\subsection{Neutron Background}

For the neutron background, we consider the two types of neutron
creation mechanisms discussed in the previous section. We have seen
that both, a radioactive neutron source and muon-induced neutrons
can be described by the same exponential energy spectrum of the
corresponding recoil events. Our model is hence 
\begin{equation} \label{eq:neutronenergyspectrum}
 \frac{\dd N_\text{n}}{\dd E}(E) =
A_\text{n}(n_\text{rad}+n_\text{muon}) \cdot \exp\left(-
\frac{E}{E_\text{dec}} \right),
\end{equation}
with $E_\text{dec}=\unit[23.54]{keV}$ as given in the previous
section.  The amplitude $A_\text{n}$ is a function of the total
expected number of neutrons, i.e.\ of the sum $n_\text{rad} +
n_\text{muon}$ of accepted single neutron scatterings due to
radioactive sources and muon-induced neutrons, respectively. Both
these numbers are unknown parameters. These parameters shall here
be defined as the \emph{sum} of expected events in \emph{all}
detector modules. Nevertheless, (\ref{eq:neutronenergyspectrum})
describes the expected energy spectrum of neutron-induced events in
each module.

Although neutrons will scatter off all three types of target nuclei
in CaWO$_4$, Monte Carlo simulations show that more than 90\% of
all scatterings detected in the energy range of interest happen off
oxygen. In our background model, we hence treat neutron events as
oxygen recoils and use the corresponding quenching factor to
describe their light yield distribution. The uncertainty of the
oxygen quenching factor as given in
Section~\ref{sec:activebackgrounddiscrimination} is included in the
likelihood, assuming a Gaussian error.   

With also other sources of events contributing to our acceptance
regions in the oxygen band, the unknown parameters $n_\text{rad}$
and $n_\text{muon}$ of the neutron background need to be
constrained by an additional reference measurement. As outlined
above, we exploit the ability to cause coincident events in more
than one module to estimate the contribution from neutrons. 

Each of the two neutron creation mechanisms considered here has a
certain expected ratio of coincident events with multiplicity $m>1$
to single scatterings:
\begin{equation}
 r_\text{rad/muon}^m := \frac{\text{exp.\ events with mult.\ $m$ in
acc.\ region}}{\text{exp. single events in acc.\ region } }
\end{equation}
The multiplicity histograms of Fig.~\ref{fig:multiplicities}
represent our measurement of these ratios. 

On the other hand, we have observed coincident events in our
background data as outlined above and we will use them to derive
constraints on the expected number of \emph{single} neutron
scatterings. To this end, we treat the occurrence of events with
each multiplicity $m$ as an independent Poisson process. This
yields a likelihood factor of the form
\begin{align}
 \mathcal{L}^m_\text{neutron}&(n_\text{rad},n_\text{muon})
=\nonumber \\* &\text{Pois}\, (\, N^m_\text{obs} \, | \, 
r^m_\text{rad} \cdot n_\text{source}   + r^m_\text{muon} \cdot
n_\text{muon} \,)
\end{align}
for each multiplicity $m$, where $\text {Pois}\,(x\,|\,y)$ denotes the
Poisson probability to observe $x$ events when $y$ are expected,
and $N^m_\text{obs}$ is the number of observed events with
multiplicity $m$. In our case, we have $N^3_\text{obs}=2$,
$N^5_\text{obs}=1$, and $N^m_\text{obs}=0$ for all other $m>1$.

Each number $N^m_\text{obs}$ can be partly due to neutrons from a
radioactive source and to muon-induced neutrons. We emphasize that,
since the multiplicity spectra for the two types of neutron
mechanisms are significantly different, a fit can distinguish the
two contributions based on the observed multiplicities. It will
ultimately choose the distribution that fits best to the data. 

For neutrons from radioactive sources, the measurement of the
coincidence rate has sufficient statistics
(Fig.~\ref{fig:multiplicities}) and we obtain the ratios
$r^m_\text{rad}$ directly by dividing the contents of the
respective bins. On the other hand, for muon-induced neutrons the
statistics of coincident events is low and statistical
uncertainties of the histogram in Fig.~\ref{fig:multiplicities}
need to be taken into account. We accomplish this by fitting the
observed multiplicity spectrum with a (purely heuristic)
exponentially decaying function. This function is then evaluated to
obtain the ratios $r^m_\text{muon}$. We directly add the likelihood
of this fit to our total likelihood function and determine the
parameters of the exponential simultaneously with all the other
estimates. This way, their statistical uncertainty is automatically
considered.

While an exponential fit to the multiplicity spectrum is clearly
not the most general form, it is evident that the abundance of
coincidences decreases towards higher multiplicities. Our fit
models this behavior and provides a reasonable approximation to the
observed spectrum, in particular at the low multiplicities which
are most relevant here.

\subsection{Lead Recoil Background}

In analogy to the $\alpha$-background, we study the lead recoil
background in a region where the Pb-band is free of other sources
of events. As discussed above, this indicates that the energy
spectrum of this background has a decreasing tail towards lower
energies which we model by an exponential starting at
\unit[90]{keV} (the upper energy bound of the lead reference
region), on top of a constant contribution:
\begin{equation} \label{eq:leadspectrum}
   \frac{\dd N_\text{Pb}}{\dd E}(E) = A_\text{Pb}^d \cdot
\left[C_\text{Pb}
+\exp\left(\frac{E-\unit[90]{keV}}{E^\text{Pb}_\text{decay}}
\right) \right] .
\end{equation}

In contrast to the simple fit of this function
discussed in
Section~\ref{sec:leadrecoils_qualitative}, we take into account the
differences between the detector modules here: The free parameter
$A_\text{Pb}^d$ may be module-dependent and describes the
different recoil background rates in the individual detector
modules, while we use the same spectral decay length
$E^\text{Pb}_\text{decay}$ and constant background term
$C_\text{Pb}$ for all modules. The latter quantities are
characteristic of the implantation profile of $\alpha$-emitters in
the clamps and can thus be assumed to be universal if the
underlying implantation mechanism is the same for all clamps.

The unknown parameters of this background model are constrained by
including the reference region in the lead band as introduced in
Section~\ref{sec:leadrecoils_qualitative}. The technical
realization is identical to the one discussed for the
$\alpha$-background. Since the other known sources of events play
only a negligible role in this region, this is purely a reference
measurement for the parameters of the lead recoil background.
Moreover, the inclusion of the reference region again allows to
treat the quenching factor of lead as a free fit parameter, which
is automatically constrained by the observed reference events. The
corresponding uncertainty is thus directly built into the
likelihood function.

\subsection{WIMP Signal}

The density $\rho_\chi(E,y \, | \, m_\chi,\sigma_\text{WN})$ of a
possible signal due to coherent WIMP-nucleon scatterings is the sum
of three components, one for each possible recoiling nucleus in
CaWO$_4$. We calculate the expected recoil energy spectrum for each
component as a function of the WIMP parameters, using the usual
standard assumption of an isothermal WIMP halo of density
\unit[0.3]{GeV/cm$^3$}, with a galactic escape velocity of
$v_\text{esc}=\unit[544]{km/s}$ and an asymptotic velocity of
$v_\infty=\unit[220]{km/s}$. Effects of the nuclear substructure
are taken into account by the Helm form factor as given in
\cite{Lewin1996_reviewmathematics}, and the resulting energy
spectra are finally convolved with the resolution of the phonon
detectors to obtain the ultimately measured spectrum. 

In the expected light yield distribution, we take into account the
uncertainties of the quenching factors for all three
nuclei, approximating the errors given in
Section~\ref{sec:activebackgrounddiscrimination} by Gaussians. We
note that the inclusion of these uncertainties has only a minor
influence onto our results and even doubling the errors has no
relevant effect. In the same way, varying the galactic escape velocity within its uncertainties is found to have negligibly
small effects.

\section{Results and Discussion} \label{sec:resultsanddiscussion}

In this section, we summarize the results obtained from the maximum
likelihood fit as introduced above. 

\subsection{Resulting Fit Parameters}

We find that the total likelihood function has two maxima in the
parameter space, which we denote M1 and M2, respectively. M1 is the
global maximum, but M2 is only slightly disfavored with respect to
M1. We will hence discuss both solutions in the following.

Table~\ref{tab:fitresults} shows the expected contributions of the
backgrounds and of a possible WIMP signal in the two likelihood
maxima. The background contributions are very similar for M1 and
M2: The expected $e/\gamma$-background is one event per module
according to the choice of the acceptance region, with a negligible
statistical uncertainty due to the large number of events in the
$e/\gamma$-band. The lead recoil and the $\alpha$-background are
similar to our simple estimates given in
Section~\ref{sec:qualitativebckdiscussion}. Both these backgrounds
are slightly larger than the contribution from neutron scatterings.
In the context of the latter, the fit assigns roughly half of the
coincident events to neutrons from a radioactive source and to
muon-induced neutrons, respectively. This translates into about
\unit[10]{\%} of the single neutron background being muon-induced.
{\setlength{\extrarowheight}{2pt}
\renewcommand{\arraystretch}{1.1}
\begin{table}
\begin{center}
\begin{tabular}{l || c | c }
 \hline
  & M1 & M2 \\ \hline
  $e/\gamma$-events & $8.00 \pm 0.05$ & $8.00 \pm 0.05$ \\   
  $\alpha$-events & $11.5\,_{-2.3}^{+2.6}$ &
$11.2\,_{-2.3}^{+2.5}$\\
  neutron events & $7.5\, _{-5.5}^{+6.3}$ & $9.7\,_{-5.1}^{+6.1}$\\
  Pb recoils & $15.0_{-5.1}^{+5.2}$ & $18.7\,_{-4.7}^{+4.9}$\\
\hline
  signal events & $29.4\,_{-7.7}^{+8.6}$ & $24.2\,_{-7.2}^{+8.1}$
\\ \hline\hline
  $m_\chi$ [GeV] & 25.3 & 11.6\\
  $\sigma_\text{WN}$ [pb] & $1.6\cdot10^{-6}$ & $3.7\cdot10^{-5}$\\
\hline
 \end{tabular}
\end{center}
 \caption{Results of the maximum likelihood fit. Shown are the
expected total contributions from the backgrounds considered as
well as from a possible WIMP signal, for the parameter values of the two likelihood maxima. The small statistical
error given for the $e/\gamma$-background reflects the large number
of observed events in the $e/\gamma$-band. The other errors
correspond to a $1\sigma$ confidence interval as determined by
MINOS (see Section~\ref{sec:likelihoodconcepts}). The
corresponding WIMP mass and interaction cross section are listed for
each of the two likelihood maxima.}
 \label{tab:fitresults}
\end{table}
}

In both likelihood maxima the largest contribution is assigned to
a possible WIMP signal. The main difference between the two
likelihood maxima concerns the best-fit WIMP mass and the
corresponding cross section, with $m_\chi=\unit[25.3]{GeV}$ in case
of M1 and $m_\chi=\unit[11.6]{GeV}$ for the case M2.
The possibility of two different solutions for the WIMP mass can be
understood as a consequence of the different nuclei present in our
target material. The given shape of the observed energy spectrum
can be explained by two sets of WIMP parameters: in the case of M1,
the WIMPs are heavy enough to detectably scatter off tungsten
nuclei (cp.\ Fig.~\ref{fig:targetnuclei}), about 69~\% of the recoils
are on tungsten, $\sim25$~\% on calcium and $\sim7$~\% on oxygen, while in
M2, oxygen (52~\%) and calcium recoils (48~\%) constitute the
observed signal and lead to a similar spectral distribution in
terms of the recoil energy. The two possibilities can, in
principle, be discriminated by the light yield distribution of the
signal events. However, at the low recoil energies in question,
there is considerable overlap between the oxygen, calcium, and
tungsten bands, so that we can currently not completely resolve the
ambiguity. This may, however, change in a future run of the
experiment.  

\begin{figure}
  \includegraphics[width=\linewidth]
{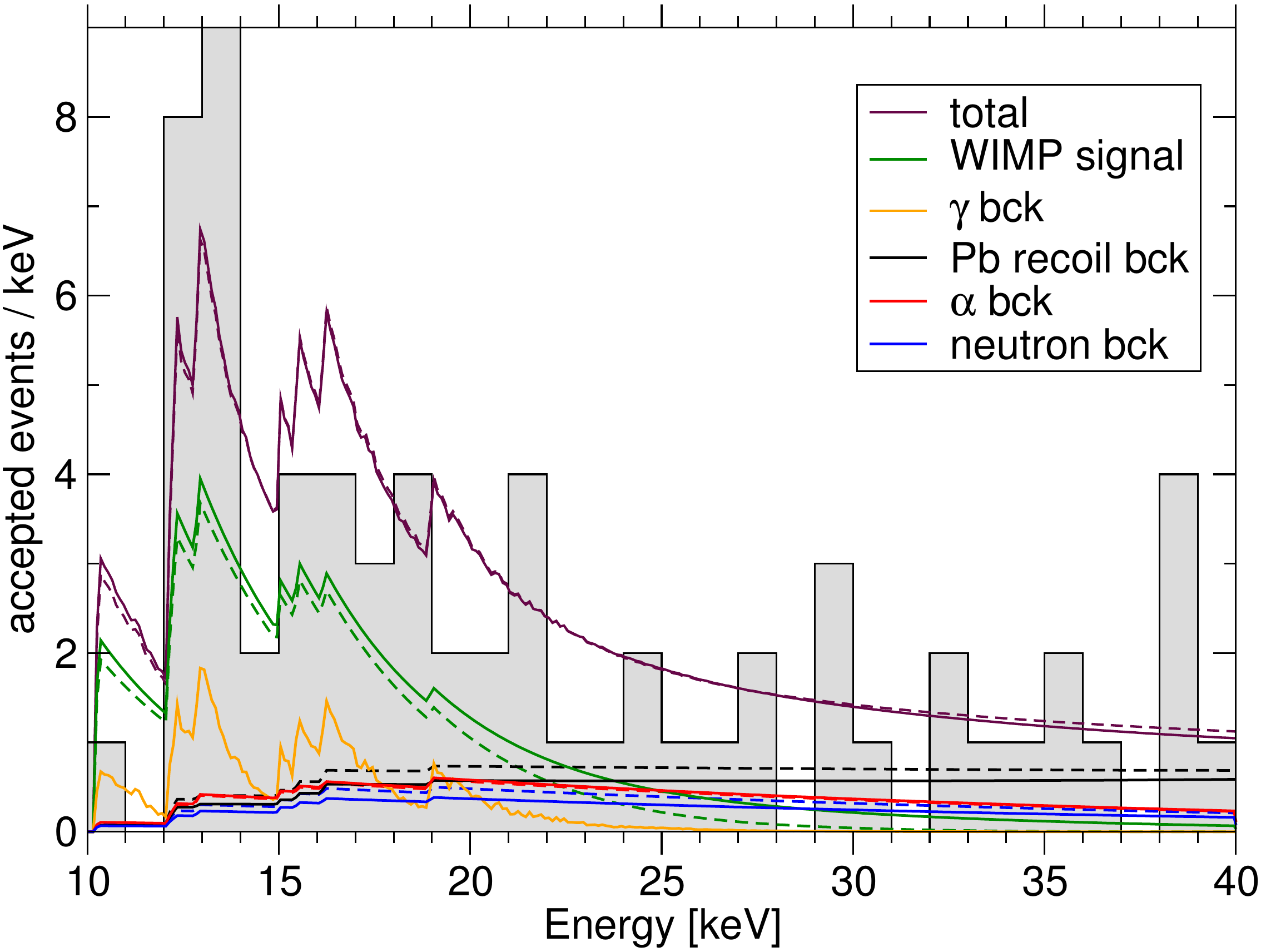}
  \caption{(Color online) Energy spectrum of the accepted events
from all detector modules, together with the expected contributions
from the considered backgrounds and a WIMP signal, as inferred from
the likelihood fit. The solid and dashed lines correspond to the
fit results M1 and M2, respectively.}
  \label{fig:fitresult_espectrum}
\end{figure}

Fig.~\ref{fig:fitresult_espectrum} illustrates the fit result,
showing an energy spectrum of all accepted events together with the
expected contributions of backgrounds and WIMP signal. The solid
lines correspond to the likelihood maximum M1, while the dashed
lines belong to M2.
The complicated shape of the expectations is the result of taking
into account the energy-dependent detector acceptances. In
particular, the different energy thresholds of the individual
detector modules lead to a steep increase of the expectations when
an additional module sets in. 

We note that neither the expected $\alpha$- or lead recoil
backgrounds nor a possible neutron background resemble a WIMP
signal in terms of the shape of their energy spectrum. Even if our
analysis severely underestimated one of these backgrounds, this
could therefore hardly be the explanation of the observed event
excess.

\begin{figure}
  \includegraphics[width=\linewidth]{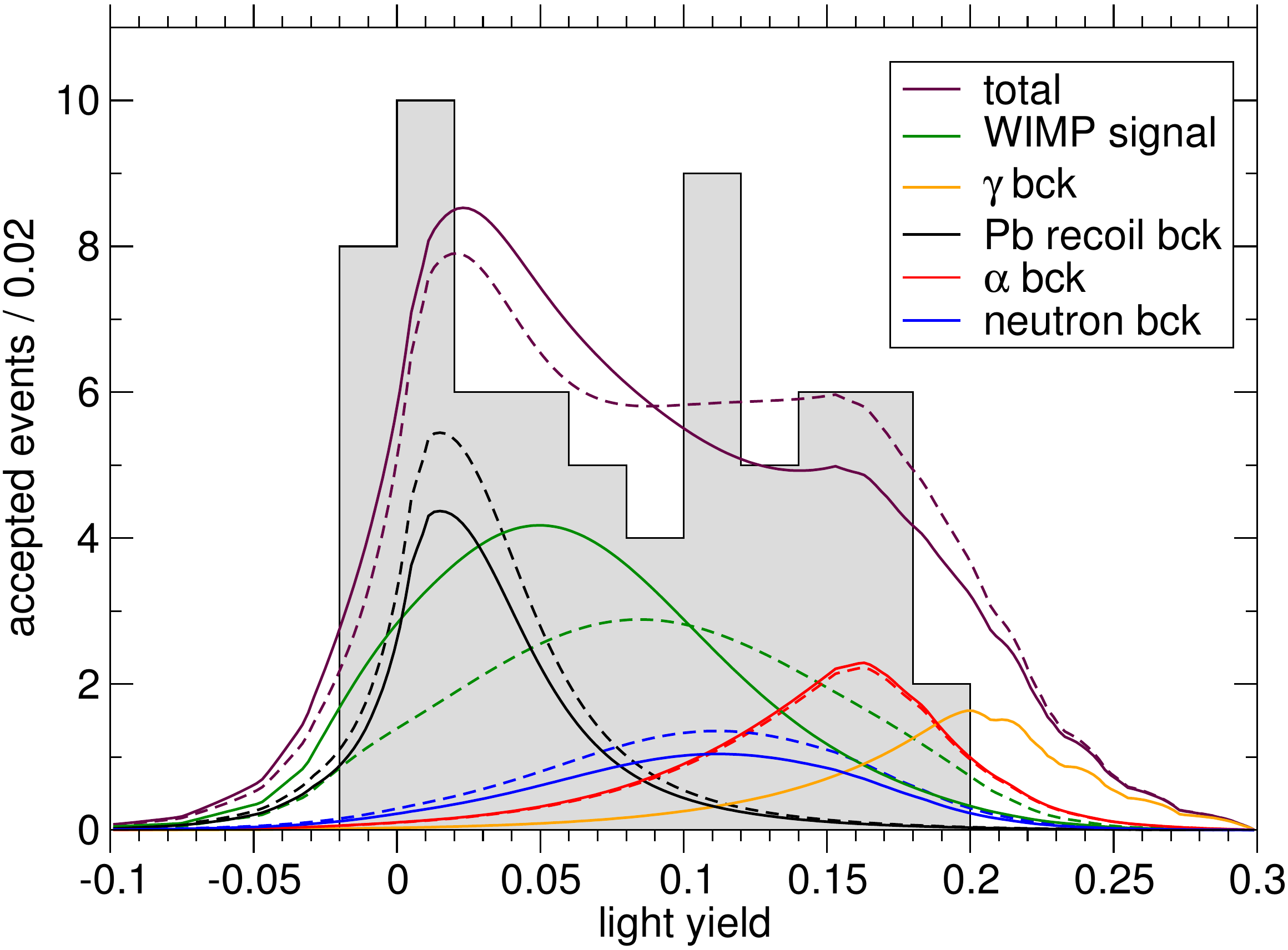}
  \caption{(Color online) Light yield distribution of the accepted
events, together with the expected contributions of the backgrounds
and the possible signal. The solid and dashed lines correspond to
the parameter values in M1 and M2, respectively.}
  \label{fig:fitresult_yspectrum}
\end{figure}

On the other hand, the leakage of $e/\gamma$-events rises steeply
towards low energies and one may be tempted to consider a strongly
underestimated $e/\gamma$-background as the source of the
observation. However, in addition to the energy spectrum, also the
distribution in the light yield parameter needs to be taken into
account. Fig.~\ref{fig:fitresult_yspectrum} shows the corresponding
light yield spectrum of the accepted events, together with the
expectations from all considered sources. 
Again, the shape of the expectations is the result of the
individual detector acceptances being considered. As expected, the
contributions from the $e/\gamma$- and also from the
$\alpha$-background quickly decrease towards lower light yields and
thus differ significantly from the expected distribution of a WIMP
signal. 

In order to check the quality of the likelihood fit, we calculate
a $p$-value according to the procedure summarized in
Section~\ref{sec:likelihoodconcepts}. We divide the energy-light
yield plane into bins of \unit[1]{keV} and 0.02, respectively, and
include the acceptance region of each module as well as the alpha-
and Pb recoil reference regions in the calculation. The two
likelihood maxima are found to give very similar results, with
$p$-values of about 0.36 and 0.35, respectively. This not very small value for $p$
indicates an acceptable description by our background-and-signal
model.

\subsection{Significance of a Signal}

As described in Section~\ref{sec:likelihoodconcepts}, the
likelihood function can be used to infer whether our observation
can be statistically explained by the assumed backgrounds alone.
To this end, we employ the likelihood ratio test. The result of
this test naturally depends on the best fit point in parameter
space, and we thus perform the test for both likelihood maxima
discussed above. The resulting statistical significances, at which
we can reject the background-only hypothesis, are
\begin{center}
\begin{tabular}{c c}
  for M1: & 4.7\,$\sigma$\\
  for M2: & 4.2\,$\sigma$.
\end{tabular}
\end{center}

In the light of this result it seems unlikely that the backgrounds
which have been considered can explain the data, and an additional
source of events is indicated.  Dark Matter particles, in the form
of coherently scattering WIMPs, would be a source with suitable
properties. We note, however, that the background contributions are still relatively large. A reduction of the overall background level will reduce remaining uncertainties in modeling these backgrounds and is planned for the next run of CRESST (see Section~\ref{sec:futuredevelopments}). 

\subsection{WIMP Parameter Space}

\begin{figure}
  \includegraphics[width=\linewidth]{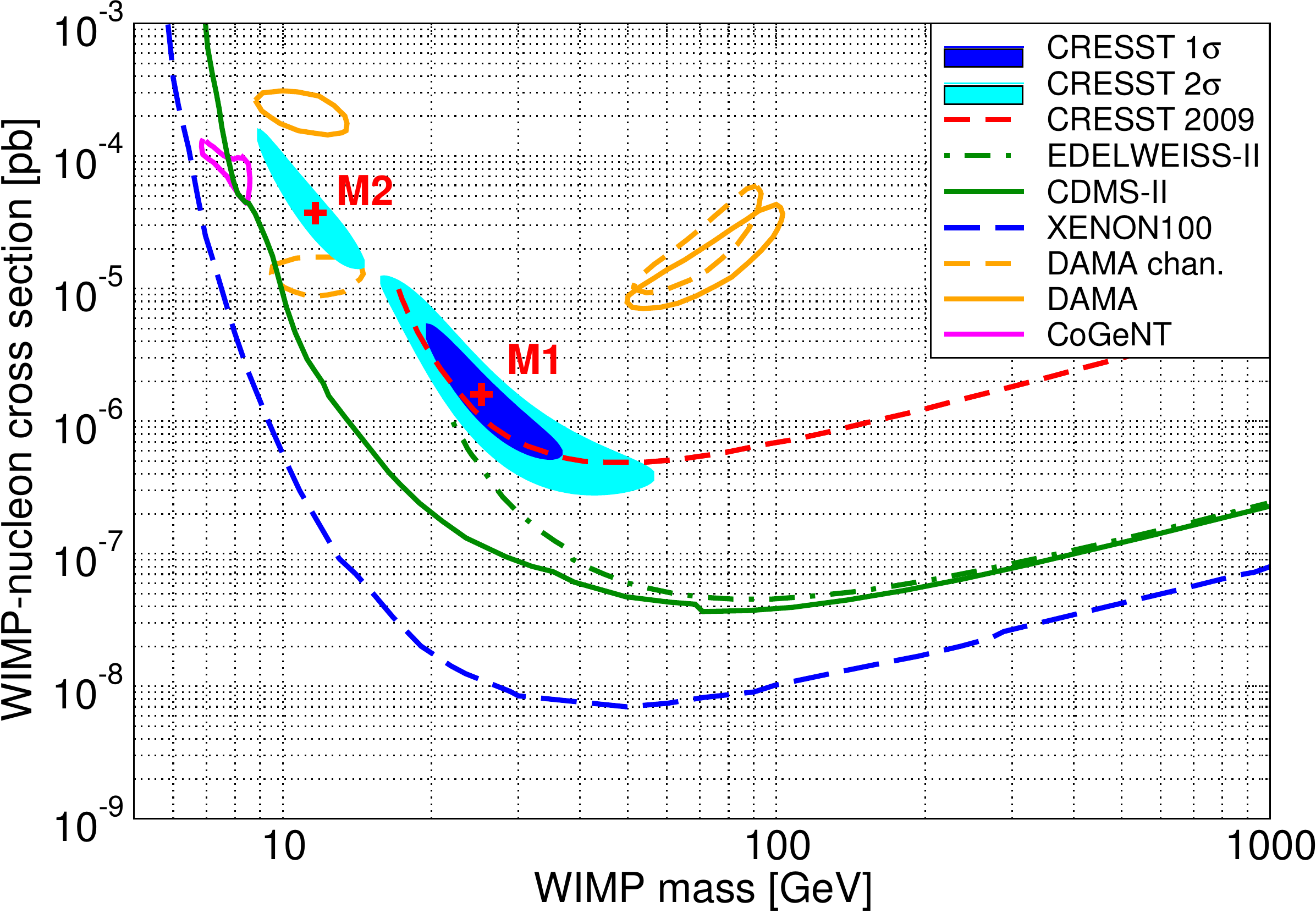}
  \caption{The WIMP parameter space compatible with the CRESST results discussed here, using the background model described in the text,
together with the exclusion limits from
\mbox{CDMS-II}~\cite{CDMS2009_FinalResults},
XENON100~\cite{XENON1002011_DMresults}, and
\mbox{EDELWEISS-II}~\cite{EDELWEISS2011_finalresults}, as well as the
CRESST limit obtained in an earlier run~\cite{Angloher2009_run30}.
Additionally, we show the 90\% confidence regions favored by
CoGeNT~\cite{Aalseth2011_CoGeNTModulation} and
DAMA/LIBRA~\cite{Savage2009_DAMAcompatibility} (without and with
ion channeling). The CRESST contours have been calculated with
respect to the global likelihood maximum M1.}
  \label{fig:msigmaplane}
\end{figure}

In spite of this uncertainty, it is interesting to study the
WIMP parameter space which would be compatible with our
observations. Fig.~\ref{fig:msigmaplane} shows the location of the
two likelihood maxima in the $(m_\chi,\sigma_\text{WN})$-plane,
together with the $1\sigma$ and $2\sigma$ confidence regions
derived as described in Section~\ref{sec:likelihoodconcepts}. The
contours have been calculated with respect to the global likelihood
maximum M1. 
We note that the parameters compatible with our observation are
consistent with the CRESST exclusion limit obtained in an earlier
run~\cite{Angloher2009_run30}, but in considerable tension with the
limits published by the \mbox{CDMS-II}~\cite{CDMS2009_FinalResults}
and \mbox{XENON100}~\cite{XENON1002011_DMresults} experiments. 
The parameter regions compatible with the
observation of DAMA/LIBRA (regions taken from
\cite{Savage2009_DAMAcompatibility}) and CoGeNT
\cite{Aalseth2011_CoGeNTModulation} are located somewhat outside
the CRESST region.

\section{Future Developments} \label{sec:futuredevelopments}
Several detector improvements aimed at a reduction of the overall background level
are currently being implemented. The most important one addresses the reduction of 
the alpha and lead recoil backgrounds. The bronze clamps holding the target crystal 
were identified as the source of these two types of backgrounds. They will be replaced 
by clamps with a substantially lower level of contamination. A significant reduction 
of this background would evidently reduce the overall uncertainties of our background 
models and allow for a much more reliable identification of the properties of a 
possible signal. 

Another modification addresses the neutron background. An additional layer of polyethylene 
shielding (PE), installed inside the vacuum can of the cryostat, will complement the 
present neutron PE shielding which is located outside the lead and copper shieldings.

The last background discussed in this work is the leakage from the $e/\gamma$-band. Most of 
these  background events are due to internal contaminations of the target crystals so that 
the search for alternative, cleaner materials and/or production procedures is of high 
importance. The material ZnWO$_4$, already tested in this run, is a promising candidate in 
this respect.

\section{Summary}

With an exposure of \unit[730]{kg\,d}, the CRESST Dark Matter search has
observed in the latest run a total of 67 events in an acceptance region of low energy
nuclear recoils. Possible background contributions to this number include
leakage of $e/\gamma$-events, events from neutrons, from alpha-particles, and from 
recoiling nuclei in $\alpha$-decays. We have estimated these four backgrounds and have
found using a likelihood ratio test that, at a significance larger than $4\sigma$,
these backgrounds are not sufficient to explain all the observed events. Scatterings
of WIMPs may be the origin of this effect and, under this assumption, we have derived 
the corresponding WIMP parameters. Finally, we have presented the plans for the next 
run of the experiment, in which we aim for a further clarification of this
hypothesis.

\section*{Acknowledgements}

This work was supported by funds of the Deutsche
For\-schungs\-gemeinschaft DFG (Transregio 27: Neutrinos and Beyond),
the Munich Cluster of Excellence (Origin and Structure of the
Universe), the Maier-Leibnitz-La\-bo\-ra\-to\-ri\-um (Garching), the Science
and Technology Facilities Council (STFC) UK, as well as the DFG
Internationales Gra\-du\-ier\-ten\-kolleg GRK 683. We are grateful to LNGS
for their generous support of CRESST, in particular to Marco Guetti
for his constant assistance. 
We would like to thank H. Kraus, S. Henry and A.~Brown for their 
numerous and vital contributions to the experiment.
We thank V. Soergel for his constant interest in all aspects of our work 
and we appreciate his strong support.

\bibliographystyle{h-physrev}
\bibliography{bibliography_Run32.bib}

\end{document}